\documentclass[aps,preprintnumbers,twocolumn,amsmath,amssymb,prb]{revtex4}
\usepackage{bbm}
\usepackage{mathrsfs}
\usepackage{amsmath}
\usepackage{graphicx}
\usepackage{epstopdf}
\usepackage{ulem}
\usepackage{hyperref}

\usepackage{color}
\newcommand{\be}{\begin{equation}}
\newcommand{\ee}{\end{equation}}
\newcommand{\bea}{\begin{eqnarray}}
\newcommand{\eea}{\end{eqnarray}}
\newcommand{\bes}{\begin{split}}
\newcommand{\ees}{\end{split}}

\begin{document}
\title{Deep Learning of Phase Transitions for Bilinear-Biquadratic Spin-1 Chains
  from Correlation Aspects}
\author{Ming-Chiang Chung$^{1}$}\email{mingchiangha@phys.nchu.edu.tw}
\author{Yu-Ting Kuo$^{1}$}
\author{Ian McCulloch$^{2}$}
\author{Yuan-Hong Tsai$^{3}$}\email{yhong.tsai@gmail.com}
\affiliation{$^1$ Physics Department, National Chung-Hsing University, Taichung, 40227, Taiwan}
\affiliation{$^2$ Physics Department, National Tsing-Hua University,
  Hsinchu, 300044, Taiwan}
\affiliation{$^3$ AI Foundation, Taipei 106, Taiwan}

\begin{abstract}
We investigate the role of different correlation functions in
reconstructing the phase diagram of the spin-1
bilinear-biquadratic chain, parametrized by the angle $\theta$
between the bilinear and biquadratic couplings, using our previously developed
label-free machine-learning pipeline \cite{Ming21, Ming23}, in which an
autoencoder compresses iDMRG correlation matrices into latent features
and K-means clustering with silhouette analysis groups them into
phases. Without any prior labels, the scheme reveals four natural
phases, the Haldane, dimerized, critical trimerized, and ferromagnetic
phases, and accurately locates the Takhtajan-Babujian ($\theta=-\pi/4$),
$SU(3)$-symmetric Uimin-Lai-Sutherland ($\theta=\pi/4$), and
Ferro-Dimer ($\theta=5\pi/4$) transition points; a shallow supervised
refinement further sharpens the phase boundaries.
We find that different correlation functions play fundamentally
different roles in phase identification. The
$SU(3)$-symmetric correlation alone selects a four-cluster partition
that can be associated with the four phases, whereas the spin-spin and dimer-dimer correlations must be
combined to recover the complete phase diagram. In contrast, the
trimer-trimer correlation behaves as a marginal correlation: it is
insufficient for phase identification by itself but improves the
precision of the predicted phase boundaries when combined with the
other correlations. Furthermore, the latent representations and
silhouette values reveal signatures that we relate to the
$\theta$ dependence of the correlations within each phase and,
tentatively, to the underlying renormalization-group structure. These
results demonstrate that machine learning can not only identify
quantum phases but also uncover the physical information encoded in
different correlation functions.
\end{abstract} 

\date{\today}
\maketitle

\section{Introduction} 
Quantum phase transitions in one-dimensional spin systems remain a
vibrant area of research, with profound implications for both
theoretical models and real materials\cite{Auerbach,  Sachdev, Giamarchi}. In particular, the spin-1
bilinear-biquadratic (BLBQ) chain provides a versatile platform to
explore a rich variety of ground states\cite{Haldane83, AKLT,
  BarberBatchelor,Uimin,Lai,Sutherland,Takhtajan,Babujian,Fath,ItoiKato,BLBQRev}, including the Haldane
phase \cite{Haldane83, AKLT}, dimerized states\cite{BarberBatchelor}, trimerized critical phases\cite{Fath,ItoiKato}, and ferromagnetic
order. These phases emerge from the competition between bilinear and
biquadratic spin interactions, parametrized by a single angle $\theta$ in the
Hamiltonian. The biquadratic spin-spin interaction, supplementing the conventional Heisenberg term, emerges as an effective low-energy interaction from higher-order superexchange processes in Hubbard-type systems\cite{MilaZhang00,Yip03}.

Among the various phases, the Haldane phase is particularly
intriguing. As the first known example of a strongly correlated
topological state\cite{Haldane83}, it is characterized by a gapped
bulk spectrum and fractional spin-$1/2$ edge states\cite{Haldane83,
  AKLT}---hallmarks of symmetry-protected topological (SPT)
order\cite{Pollmann10,Pollmann12}. Consequently, the BLBQ model
exhibits topological phase transitions (TPTs), which are fundamentally
distinct from conventional transitions. Unlike symmetry-breaking
transitions, TPTs occur without any local order parameter or broken
symmetry, making them inherently difficult to detect using traditional
diagnostic tools\cite{Pollmann12, WenBook}.

Experimentally, various quasi-one-dimensional spin-1 systems have been
studied as realizations of the BLBQ model. Neutron scattering
experiments on compounds such as CsNiCl$_3$ and Y$_2$BaNiO$_5$ have confirmed
key features of the Haldane phase, including the presence of an
excitation gap and multiparticle continua.~\cite{Buyers1986,Morra1988,
Xu1996,Kenzelmann2002}
Other materials, like
LiVGe$_2$O$_6$, exhibit spin-Peierls transitions potentially driven by
strong biquadratic interactions, highlighting the role of lattice
effects and spin-phonon coupling.~\cite{Millet1999}
In addition, several organic
coordination compounds, such as NENP, NINO, and NDMAP, have been
synthesized to realize antiferromagnetic spin-1 chains with high
structural tunability and well-isolated one-dimensional
behavior.~\cite{BLBQRev}
These systems offer clean experimental platforms to
investigate topological phases and quantum criticality in low
dimensions.

Several well-established numerical techniques such as
exact diagonalization, quantum Monte Carlo simulations,
and the density matrix renormalization group (DMRG)
have been widely employed to compute correlation
functions and ground-state wave functions in quantum
many-body systems.~\cite{Dagotto1994,Sandvik2010,White1992,White1993,Schollwock2011}
While these methods have provided valuable insights into
the BLBQ model, obtaining a complete phase diagram
remains computationally demanding because the ground
state and correlation functions must be recalculated for
each value of the interaction parameter.~\cite{Botet1983,Nightingale1986,Nomura1989,Sakai1990,WhiteHuse1993,SorensenAffleck1993,Schollwock1995,Schollwock1996,Lauchli2006}
This challenge becomes particularly significant when
high-resolution scans are required to accurately determine
phase boundaries or resolve narrow critical regions.

In recent years, machine learning (ML) has emerged as a promising
alternative, offering the potential to uncover hidden structures in
data such as correlation patterns, entanglement features, or wave
function amplitudes.~\cite{LeCun2015,Carleo2019} Unlike traditional approaches that rely on
predefined observables or theoretical assumptions, ML methods are
inherently data-driven. Neural networks, for instance, can be trained
on raw input data to extract compressed representations~\cite{Hinton2006} or assign
phase labels, and can generalize to classify previously unseen
configurations. This flexibility has sparked considerable interest in applying
ML techniques to detect quantum phase transitions and
identify phase boundaries in complex systems, including
those where conventional order parameters are difficult
to define.~\cite{Wang2016,Carrasquilla2017,vanNieuwenburg2017,Wetzel2017,ZhangKim2017,ZhangShenZhai2018,RodriguezNieva2019,ScheurerSlager2020,Ming20,Ming21}

Among machine learning strategies applied to phase
classification, two main approaches have emerged: supervised and unsupervised learning.~\cite{Carrasquilla2017,
vanNieuwenburg2017,Broecker2017,Chng2017,Beach2018,Ming20,Ming21}
Supervised methods
rely on labeled data, associating each configuration with
a known phase, and are typically effective at sharply resolving phase boundaries.~\cite{Carrasquilla2017,
vanNieuwenburg2017,Ming20}
However, their use depends on
prior knowledge of the system, such as the number or type
of phases, which limits their applicability in exploratory
contexts.

Unsupervised learning, by contrast, infers structure directly from unlabeled data, making it well suited for discovering unknown phases or transitions.~\cite{Wang2016,
Wetzel2017,Hu2017,RodriguezNieva2019,Ming21}
Although more
challenging, this approach enables the analysis of raw
data without assuming predefined labels or features. In
our earlier work, we introduced a hybrid protocol that
begins with unsupervised learning to reveal candidate
phase structures.~\cite{Ming21}
The process starts with an autoencoder (AE) that compresses correlation data into a compact latent representation.~\cite{Hinton2006,Ming21}
Dimensionality reduction via
principal component analysis (PCA) is then applied, followed by clustering using the K-means algorithm.~\cite{MacQueen1967}
To estimate the number of distinct phases, we evaluate the
quality of clustering using silhouette scores.~\cite{Rousseeuw1987}
Finally, the most confidently identified points are used to initialize a supervised classifier, refining the phase boundaries
with greater accuracy.~\cite{Ming21}
This method has proven effective,
particularly in detecting topological phase transitions using quantum-information-based observables as inputs.
~\cite{Ming21}

For magnetic systems, spin-spin interactions play a
central role. In our previous work, we suggested that
spin-spin correlations can be used to identify different
phases.~\cite{Ming23}
For the XY and XXZ models, we found that appropriate spin-spin correlations suffice to determine the
phase boundaries.~\cite{Ming23}
However, in the BLBQ model, a richer
variety of phases emerges, requiring more complex and
diverse correlation functions as inputs to precisely locate
the phase boundaries.

In the ferromagnetic phase, spin-spin correlations remain the most
relevant. In contrast, antiferromagnetic phases allow various ways for
spins to form singlets. When two spins form a singlet, the system
enters a dimerized phase, where dimer-dimer correlations are
essential. When three spins form a singlet, the system exhibits a
trimerized phase, for which trimer-trimer correlations are
necessary. Another notable case arises when the spin-1 on each site is
collectively separated into two spin-$1/2$, and the spin-$1/2$ of
neighboring sites form singlets\cite{AKLT}. This configuration gives
rise to a topological state known as the Haldane phase~\cite{AffleckHaldane87}, typically
identified using the correlation functions of the non-local string
order parameters ~\cite{denNijs1989, KennedyTasaki1992, Pollmann10,Pollmann12}.

Nonlocal string correlations provide a useful diagnostic of the
Haldane phase, but conventional string order does not furnish a
universal characterization of symmetry-protected topological
phases. Its interpretation depends on the symmetries of the state and
the choice of string and endpoint operators. Here we investigate
whether the phase structure of the BLBQ chain can be recovered using
only finite-range correlations of local operators. We therefore omit
string correlators from the machine-learning inputs and examine the
complementary information contained in the spin-spin, dimer-dimer,
trimer-trimer, and $SU(3)$ correlation channels. In particular, the
$SU(3)$-symmetric correlation, motivated by the $SU(3)$-symmetric
Uimin-Lai-Sutherland (ULS) point~\cite{Uimin,Lai,Sutherland}, turns
out to suffice for identifying the Haldane phase alongside the other
phases.

Although machine learning has been widely applied to identify quantum
phases, a fundamental question remains largely unexplored: what
physical information is actually encoded in different correlation
functions, and how does this information affect phase identification?
The spin-1 BLBQ chain provides an ideal
platform to address this question because its phase diagram contains
topological, symmetry-emergent, symmetry-breaking, and first-order
phases simultaneously. By systematically examining different
correlation functions, one can not only evaluate their effectiveness
for phase identification but also gain insight into the physical
information that they encode.

Sec.~\ref{sec:BLBQ} introduces the BLBQ model and the related phases.
Sec.~\ref{sec:unsupervised} describes the unsupervised method and the
pipeline used in this work. Sec.~\ref{sec:results} contains the
results obtained using the ML pipeline. Sec.~\ref{sec:discussion}
discusses the relation between the ML results, critical phenomena, and
the renormalization-group analysis. Finally,
Sec.~\ref{sec:conclusion} summarizes the results and provides an
outlook.

\section{BLBQ model} 
\label{sec:BLBQ}

\subsection{Method and phase diagram}

The Hamiltonian of the one-dimensional spin-$1$ BLBQ model is given by
\begin{equation}
H =
\sum_i
\left[
\cos\theta \, \mathbf{S}_i \cdot \mathbf{S}_{i+1}
+
\sin\theta \left( \mathbf{S}_i \cdot \mathbf{S}_{i+1} \right)^2
\right],
\end{equation}
where $\mathbf{S}_i$ denotes the spin-$1$ operator at site $i$, and
$\theta \in [0,2\pi)$, equivalently $\theta \in [-\pi,\pi)$, controls
the relative strength of the bilinear and biquadratic interactions.

\begin{figure}[th]
  \begin{center}
    \includegraphics[width=9cm]{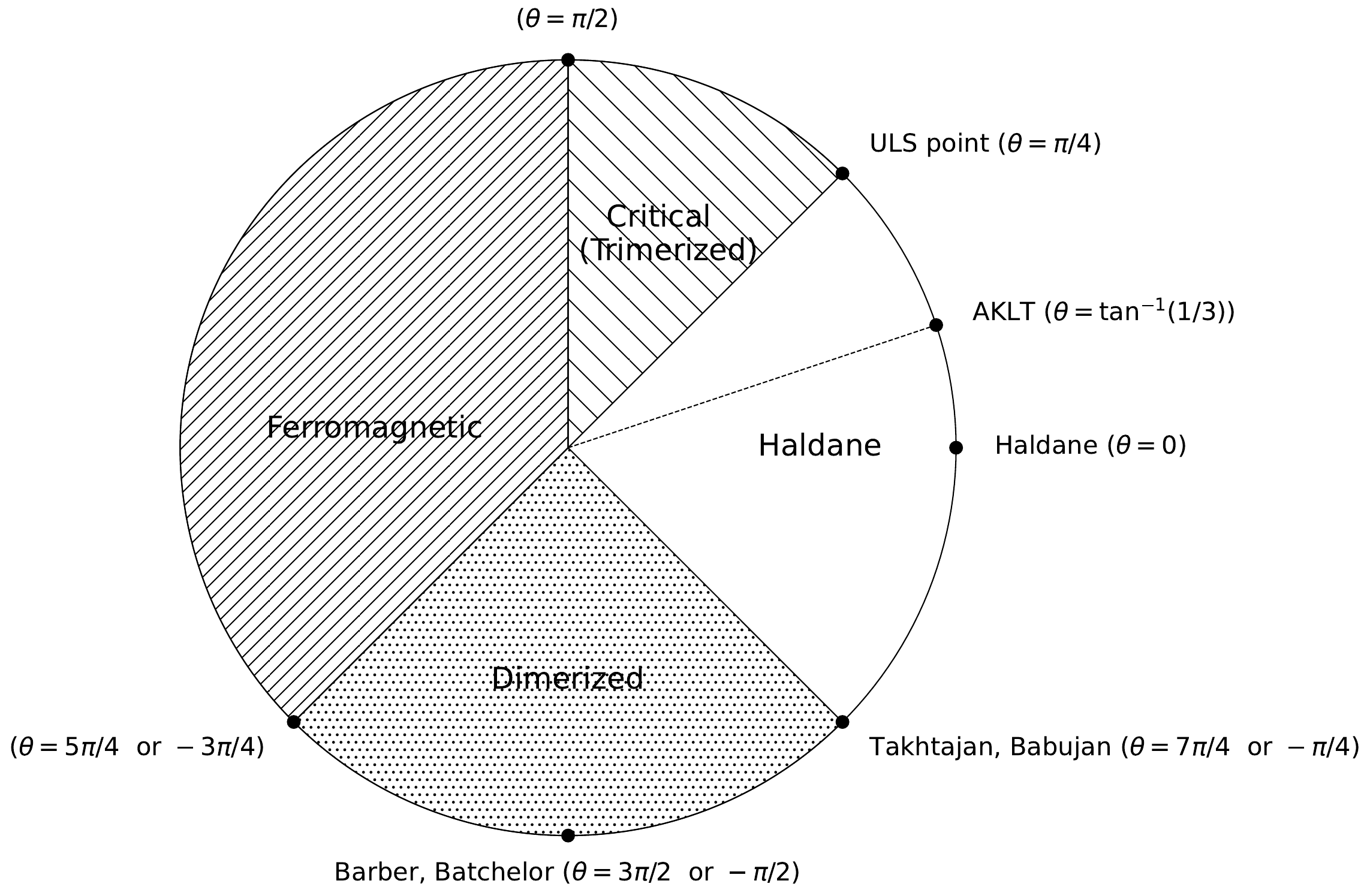}
    \caption{Phase diagram of the BLBQ model.}
    \label{fig:PD}
  \end{center}
\end{figure}

Fig~\ref{fig:PD} shows the phase diagram of the BLBQ model. For
$\pi/2 < \theta < 5\pi/4$, the ground state is ferromagnetic (FM). In
the interval $5\pi/4 < \theta < 7\pi/4$, equivalently
$-3\pi/4 < \theta < -\pi/4$, the system enters the dimerized phase
(DM). For $-\pi/4 < \theta < \pi/4$, it realizes the Haldane phase
(HP). Finally, in the region $\pi/4 < \theta < \pi/2$, the system
exhibits a critical trimerized phase, with divergent correlation
lengths throughout.

In contrast to the spin-$1/2$ XXZ chain, which is exactly solvable by
the Bethe Ansatz for all anisotropies, only a few special points of the
BLBQ model are exactly solvable. At $\theta=3\pi/2$, the model reduces
to the antiferromagnetic biquadratic chain solved by Barber and
Batchelor~\cite{BarberBatchelor}. Its ground state is two-fold
degenerate and separated from the excitations by a finite gap,
corresponding to the dimerized phase. At $\theta=-\pi/4$, the exact
solution was obtained by Takhtajan~\cite{Takhtajan} and
Babujian~\cite{Babujian} using the Bethe Ansatz. At $\theta=\pi/4$, the
Hamiltonian becomes equivalent to the three-component lattice gas with
$SU(3)$ symmetry, solved by Uimin~\cite{Uimin}, Lai~\cite{Lai}, and
Sutherland~\cite{Sutherland}. This special point is known as the
Uimin-Lai-Sutherland (ULS) point. Meanwhile, at
$\theta=\tan^{-1}(1/3)$, Affleck, Kennedy, Lieb, and Tasaki
(AKLT)~\cite{AKLT} constructed the exact singlet ground state, the
valence-bond-solid (VBS) state, which provides a paradigmatic
realization of the Haldane phase.

Historically, the spin-$1$ Haldane phase was first identified by
Haldane, who analyzed the spin-$1$ antiferromagnetic Heisenberg chain
($\theta=0$) using the nonlinear sigma model within the framework of
quantum field theory~\cite{Haldane83, HaldanePLA83}. In contrast to the spin-$1/2$ antiferromagnetic
Heisenberg chain, which is gapless and exhibits algebraically decaying
spin-spin correlations, the spin-$1$ chain is gapped, with
correlations that decay exponentially. Haldane further conjectured
that these results extend more generally: half-integer
antiferromagnetic spin chains are gapless, whereas integer-spin chains
are gapped.

Since the BLBQ model is not exactly solvable in general, numerical
approaches are required to determine its ground states and
correlations. Early studies employed exact diagonalization using the
Lanczos method, which captured the Haldane gap, albeit with
finite-size effects.~\cite{Botet1983,Nightingale1986,Nomura1989,Sakai1990}
More accurate results for larger system sizes were later obtained
through quantum Monte Carlo simulations.~\cite{Dagotto1994,Sandvik2010}
With the development of density matrix renormalization group (DMRG)
methods by White and collaborators, highly precise calculations became
possible, allowing reliable determination of correlations.~\cite{White1992,White1993,WhiteHuse1993,SorensenAffleck1993,Schollwock1995,Schollwock1996,Schollwock2011}

\subsubsection{$\pi/2 < \theta < 5\pi/4$: Ferromagnetic phase}

In the ferromagnetic phase, the ground states of a finite chain of
$L$ spin-$1$ sites form the maximum-spin multiplet
($S_{\mathrm{tot}}=L$). Fully polarized product states are exact
ground states and represent the spontaneously broken $SU(2)$ symmetry
in the thermodynamic limit. This zero-temperature order is compatible
with the Mermin-Wagner theorem~\cite{Mermin1966} because the
low-energy magnons have quadratic rather than linear dispersion
(see also Ref.~\cite{PitaevskiiStringari1991}).

The absence of a symmetry-breaking field does not require the ground
state to be $SU(2)$ invariant: the maximum-spin multiplet contains no
invariant pure state. A rotationally invariant mixed state can instead
be constructed by averaging polarized states over their
orientations~\cite{Auerbach}. This average has vanishing local
magnetization but retains the exact ferromagnetic scalar correlations,
\be
  C_{\mathrm{SS}}(r)
  =
  \langle \mathbf{S}_i \cdot \mathbf{S}_{i+r} \rangle
  = 1 , \qquad r>0 .
\ee
These correlations therefore identify ferromagnetic alignment without
selecting a magnetization direction, and the spin-spin correlation is
the relevant correlation for this phase; its long-range order is
clearly visible in Fig.~\ref{fig:CF_FMDM}(c).

The two boundaries of the ferromagnetic phase are of different
character. At $\theta=\pi/2$ the Hamiltonian reduces to the purely
biquadratic model, which possesses an $SU(3)$ symmetry with conjugate
representations on alternating sites and a macroscopically degenerate
ground-state manifold; the transition into the ferromagnetic phase is
conventionally regarded as first order~\cite{Fath,Lauchli2006}. At
$\theta=5\pi/4$ the model becomes the $SU(3)$ ferromagnet, and the
transition into the dimerized phase is direct but not a conventional
first-order transition: the ground states at this point are highly
degenerate and scale but not conformally invariant, and the
dimerization sets in extremely slowly on the dimerized
side~\cite{Buchta2005,Lauchli2006,Dai2022}.

\subsubsection{$5\pi/4 < \theta < 7\pi/4$: Dimerized phase}

In this regime, neighboring spins tend to form singlet pairs. A single
dimer covering breaks translational symmetry by selecting either the
even or the odd bonds. In the thermodynamic limit, the two dimerized
ground states become degenerate and are related by a one-site
translation. The dimerized phase is therefore characterized by
spontaneous breaking of translational symmetry and alternating
strong- and weak-bond structures.

The spin-spin correlation function alone does not provide a
distinctive characterization of the dimerized phase. Although the
exponential decay of spin-spin correlations indicates that the system
is gapped, similar behavior also appears in the Haldane phase.
Therefore, spin-spin correlations alone are insufficient to
distinguish these phases within machine-learning analysis.

\begin{figure}[th]
	\begin{center}
	\includegraphics[width=9cm]{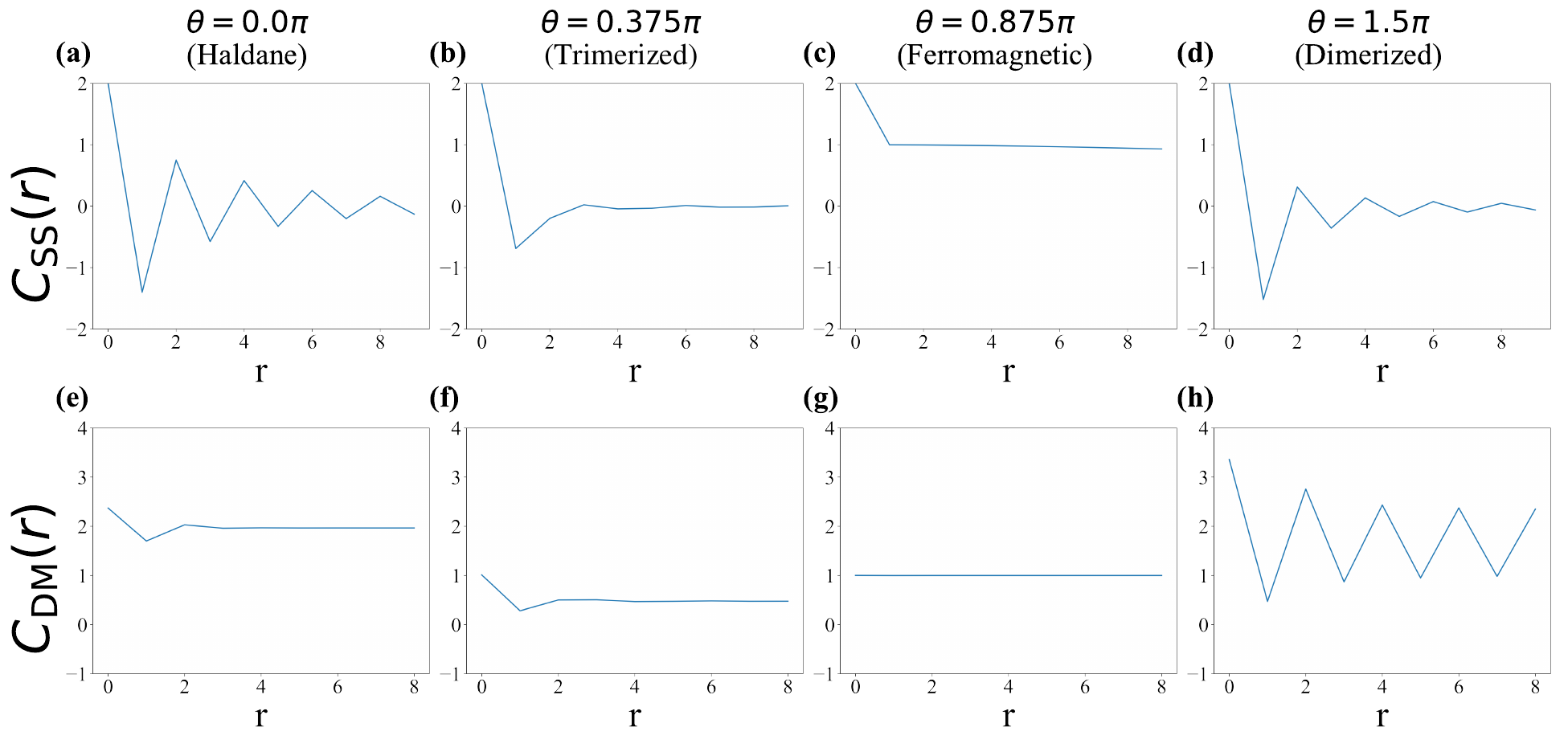}
		\caption{Spin-spin correlations $C_{\mathrm{SS}}(r)$ (upper row) and
dimer-dimer correlations $C_{\mathrm{DM}}(r)$ (lower row) at
representative points in the Haldane ($\theta=0$), critical
trimerized ($\theta=0.375\pi$), ferromagnetic ($\theta=0.875\pi$),
and dimerized ($\theta=1.5\pi$) phases.}
		\label{fig:CF_FMDM}
	\end{center}
\end{figure}

The relevant correlations are the dimer-dimer correlations, defined as
\be
  C_{\mathrm{DM}}(r) = \langle D_i D_{i+r}\rangle ,
\ee
where
\be
  D_i = \mathbf{S}_i \cdot \mathbf{S}_{i+1}.
\ee
Since $\langle D_i\rangle\neq0$ in every phase, $C_{\mathrm{DM}}(r)$
has a background at all distances; dimer order is signaled by its
alternating, period-two component, which persists to large $r$ in the
dimerized phase and provides the most relevant information for
identifying this phase. The dimer-dimer correlations in the four phases are shown
in Fig.~\ref{fig:CF_FMDM}(e)-(h).

At the special point $\theta=3\pi/2$, equivalently
$\theta=-\pi/2$, the model reduces to the purely biquadratic
Hamiltonian, which was solved exactly by Barber and Batchelor~\cite{BarberBatchelor}. The
exact solution confirms that the ground state is doubly degenerate in
the thermodynamic limit and is separated from the excited states by a finite gap.

At $\theta=7\pi/4$, equivalently $\theta=-\pi/4$, the model is at the
boundary between the dimerized and Haldane phases and was solved
exactly by Takhtajan and Babujian using the Bethe
Ansatz~\cite{Takhtajan,Babujian}. From the
conformal-field-theory perspective, this critical point is described
by the $SU(2)_2$ Wess-Zumino-Witten model~\cite{Nomura99}. As pointed out by Affleck
and Haldane~\cite{AffleckHaldane87}, the $SU(2)_2$ fixed point admits
a relevant, symmetry-allowed perturbation, whose coupling is set by
the deviation $\theta+\pi/4$ from the Takhtajan-Babujian point. This
perturbation opens a gap for either sign of the coupling, but the
resulting phases differ. For one sign the WZW field $g$ is driven
toward $g=\pm 1$, and the two solutions correspond to the two dimer
coverings related by a one-site translation. The resulting massive
phase is therefore spontaneously dimerized, which provides a
field-theoretical interpretation of why the phase on this side of the
Takhtajan-Babujian point is dimerized.

\begin{figure}[th]
	\begin{center}
	\includegraphics[width=9cm]{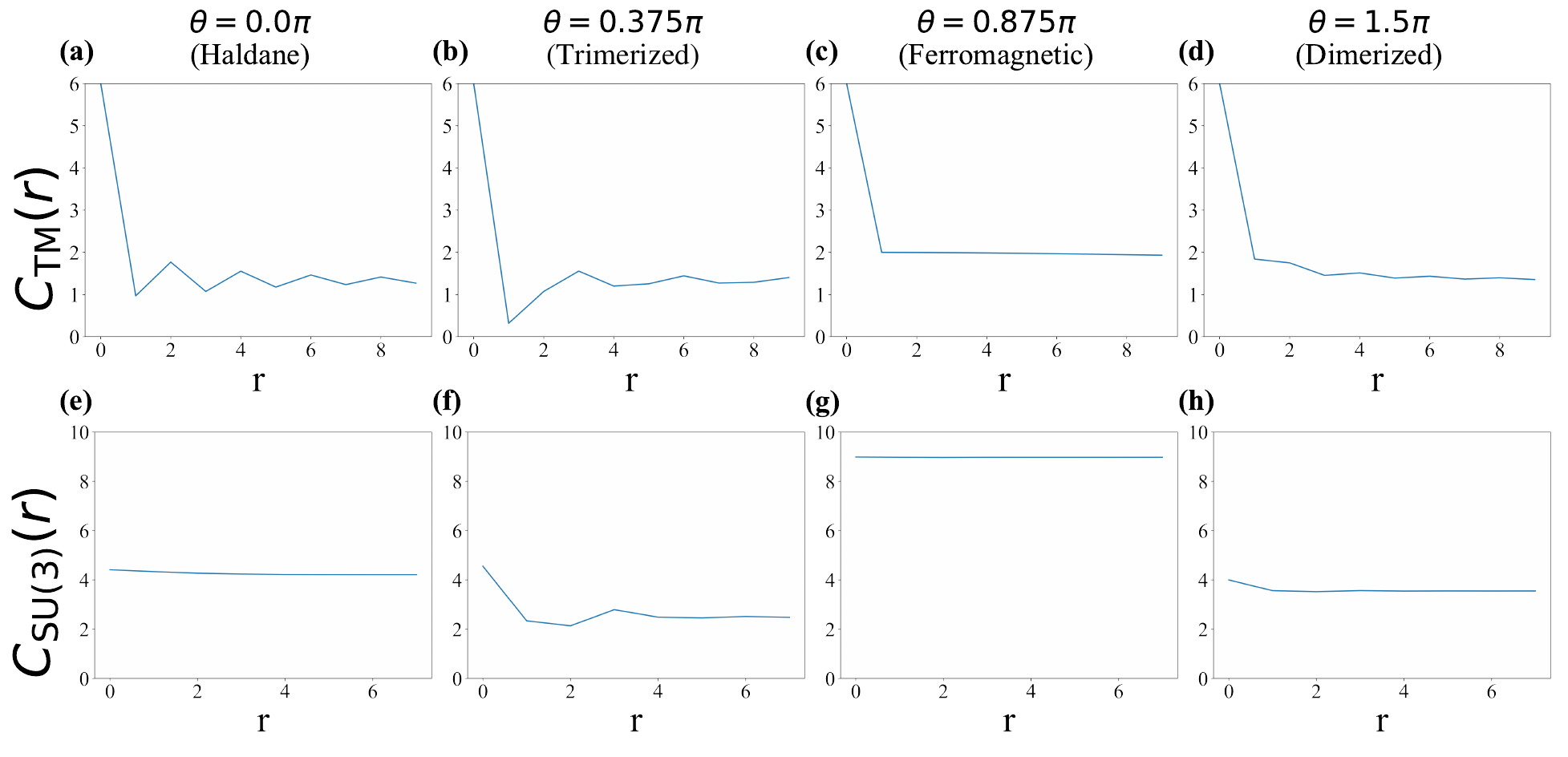}
		\caption{Trimer-trimer correlations $C_{\mathrm{TM}}(r)$ (upper row) and
$SU(3)$ correlations $C_{\mathrm{SU(3)}}(r)$ (lower row) at the same
representative points as in Fig.~\ref{fig:CF_FMDM}.}
		\label{fig:CF_TMSU3}
	\end{center}
\end{figure}

\subsubsection{$-\pi/4 < \theta < \pi/4$: Haldane phase}

For the opposite sign of the coupling, the scenario changes. The WZW
field $g$ is now driven toward the traceless configurations
$g = i\,\mathbf{n}\cdot\boldsymbol{\sigma}$, where $\mathbf{n}$ is a
unit vector: the longitudinal component $\mathrm{tr}\,g$ becomes
massive, while the transverse fluctuations of $\mathbf{n}$ are
described by the $O(3)$ nonlinear $\sigma$ model, as in Haldane's
original treatment of the spin-$1$ Heisenberg chain. The Wess-Zumino
term reduces to the topological $\Theta$-term of the $O(3)$ model with
$\Theta = k\pi = 2\pi$, in agreement with Haldane's $\Theta=2\pi S$
for $S=1$. Since the topological charge $Q\in\pi_2(S^2)=\mathbb{Z}$
is an integer, this term is trivial, and the $O(3)$ nonlinear $\sigma$
model without a topological term is disordered and gapped. The
rotational symmetry thus remains unbroken, and the system enters the
gapped Haldane phase~\cite{AffleckHaldane87}.

The Haldane phase is characterized by exponentially decaying
spin-spin correlations and fractionalized spin-$1/2$ edge states
under open boundary conditions. Its hidden topological structure is
conventionally detected by the nonlocal string order parameter~\cite{denNijs1989, KennedyTasaki1992,Pollmann10,Pollmann12},
\be
  \mathcal{O}_{\mathrm{string}}^{z}(r)
  =
  \lim_{|i-j|\to\infty}
  \left\langle
  S_i^{z}
  \exp\left(
  i\pi \sum_{k=i+1}^{j-1} S_k^{z}
  \right)
  S_{i+r}^{z}.
  \right\rangle ,
\ee

Although the string order parameter provides a standard
characterization of the Haldane phase, the corresponding nonlocal
correlations are difficult to access experimentally. In this work, we
therefore do not use string correlations as machine-learning inputs.
Instead, we focus on local correlation functions that are more
directly related to experimentally accessible observables while still
carrying sufficient information to distinguish the Haldane phase from
the neighboring phases.

At the special point $\theta=\tan^{-1}(1/3)$, Affleck, Kennedy, Lieb,
and Tasaki constructed the exact ground state of the BLBQ model, now
known as the AKLT state. In this picture, each spin-$1$ degree of
freedom is decomposed into two virtual spin-$1/2$ objects, and
neighboring virtual spins form singlet bonds. The resulting ground
state is a valence-bond-solid state with a finite excitation gap and
exponentially decaying spin-spin correlations\cite{AKLT, AKLTCMP}.
The AKLT state thus provides an explicit realization of the Haldane
phase: the unpaired virtual spin-$1/2$ objects at the ends of an open
chain are precisely the fractionalized edge states mentioned above.

\subsubsection{$\pi/4 < \theta < \pi/2$: Critical trimerized phase}

At $\theta=\pi/4$, the BLBQ model reaches the
Uimin-Lai-Sutherland (ULS) point \cite{Uimin,Lai,Sutherland}. At this point, the Hamiltonian
acquires an enlarged $SU(3)$ symmetry and becomes exactly solvable by
the Bethe Ansatz. In the continuum limit, the corresponding low-energy
theory is described by the $SU(3)_1$ Wess-Zumino-Witten model\cite{Affleck88}.

The conformal analysis of the ULS point was given by Itoi and Kato~\cite{ItoiKato}.
Their renormalization-group analysis of the marginal interaction shows
that, for $\theta>\pi/4$, the RG flow is toward the gapless
$SU(3)_1$ critical point, whereas for $\theta<\pi/4$, the perturbation
drives the system toward a massive phase, with a gap that opens
exponentially. The transition at the ULS point is therefore of the
Berezinskii-Kosterlitz-Thouless type rather than second order. This field-theoretical
picture explains why the phase considered here is critical, while the
opposite direction opens the Haldane gap, as discussed above.

Although this regime remains critical, it also exhibits strong
trimerized characteristics inherited from the nearby
$SU(3)$-symmetric structure. To capture this feature, we introduce the
trimer operator~\cite{Greiter2007}
\be
  T_i =
  \mathbf{S}_i \cdot \mathbf{S}_{i+1}
  + \mathbf{S}_{i+1} \cdot \mathbf{S}_{i+2}
  + \mathbf{S}_{i} \cdot \mathbf{S}_{i+2},
\ee
and the corresponding trimer-trimer correlation function
\be
  C_{\mathrm{TM}}(r) = \langle T_i T_{i+r}\rangle .
\ee

The dominant correlations show pronounced three-site patterns,
reflecting the tendency toward trimer formation. Therefore, while true
long-range trimer order is absent due to criticality, trimerized
correlations still provide the most relevant signatures of this phase.
The period-$3$ behavior of these correlations is shown in
Fig.~\ref{fig:CF_TMSU3}(a)-(d).

Since the RG flow is toward the $SU(3)_1$ critical point, another
useful observable is the $SU(3)$ correlation function, defined as
\be
  C_{\mathrm{SU(3)}}(r)
  =
  \left\langle
  \mathbf{S}_i \cdot \mathbf{S}_{i+r}
  +
  \left(\mathbf{S}_i \cdot \mathbf{S}_{i+r}\right)^2
  \right\rangle .
\ee
The $SU(3)$ correlations in the four phases are shown in
Fig.~\ref{fig:CF_TMSU3}(e)-(h). In contrast to the other phases, where the
$SU(3)$ correlations are nearly flat, the critical trimerized phase
exhibits a clear period-$3$ structure.

\subsection{Correlation functions}

To characterize the different phases of the BLBQ model, we employ
several correlation functions motivated by the corresponding symmetry
and field-theoretical structures. These include the spin-spin,
dimer-dimer, trimer-trimer, and $SU(3)$ correlation functions
introduced above. The corresponding correlation functions are shown in
Figs.~\ref{fig:CF_FMDM} and~\ref{fig:CF_TMSU3}.

For the machine-learning analysis, each correlation function is
arranged as a matrix rather than as a function of the distance alone.
For a block of $l_s$ consecutive sites, we evaluate the two-site
correlation $C_{\mathrm{X}}(i,j)$ between every pair of sites $i,j$
within the block, with
$\mathrm{X}\in\{\mathrm{SS},\mathrm{DM},\mathrm{TM},SU(3)\}$, so that
the curves $C_{\mathrm{X}}(r)$ shown in Figs.~\ref{fig:CF_FMDM}
and~\ref{fig:CF_TMSU3} correspond to the entries with $|i-j|=r$. Each
$l_s\times l_s$ matrix (for the multi-site operators $D_i$ and $T_i$, see
Sec.~\ref{sec:data}) is treated as a single-channel image whose
entries play the role of pixel values, and the matrices of the
correlation functions included in a given analysis are stacked as
separate channels of one input image, following
Refs.~\cite{Ming21,Ming23}. In the following we refer to these inputs
as correlation matrices.

\section{Unsupervised learning}
\label{sec:unsupervised}

\subsection{Machine-learning protocol}
\label{sec:protocol}

The goal of the machine-learning analysis is to determine whether the
phase structure of the BLBQ model can be recovered directly from the
correlation data, without assigning phase labels in advance. The input
data are the correlation matrices constructed from the spin-spin,
dimer-dimer, trimer-trimer, and $SU(3)$ correlation functions
introduced in the previous section. These matrices contain different
aspects of the many-body correlations and are used as the raw
information supplied to the learning algorithm.

\begin{figure}[th]
  \begin{center}
    \includegraphics[width=9cm]{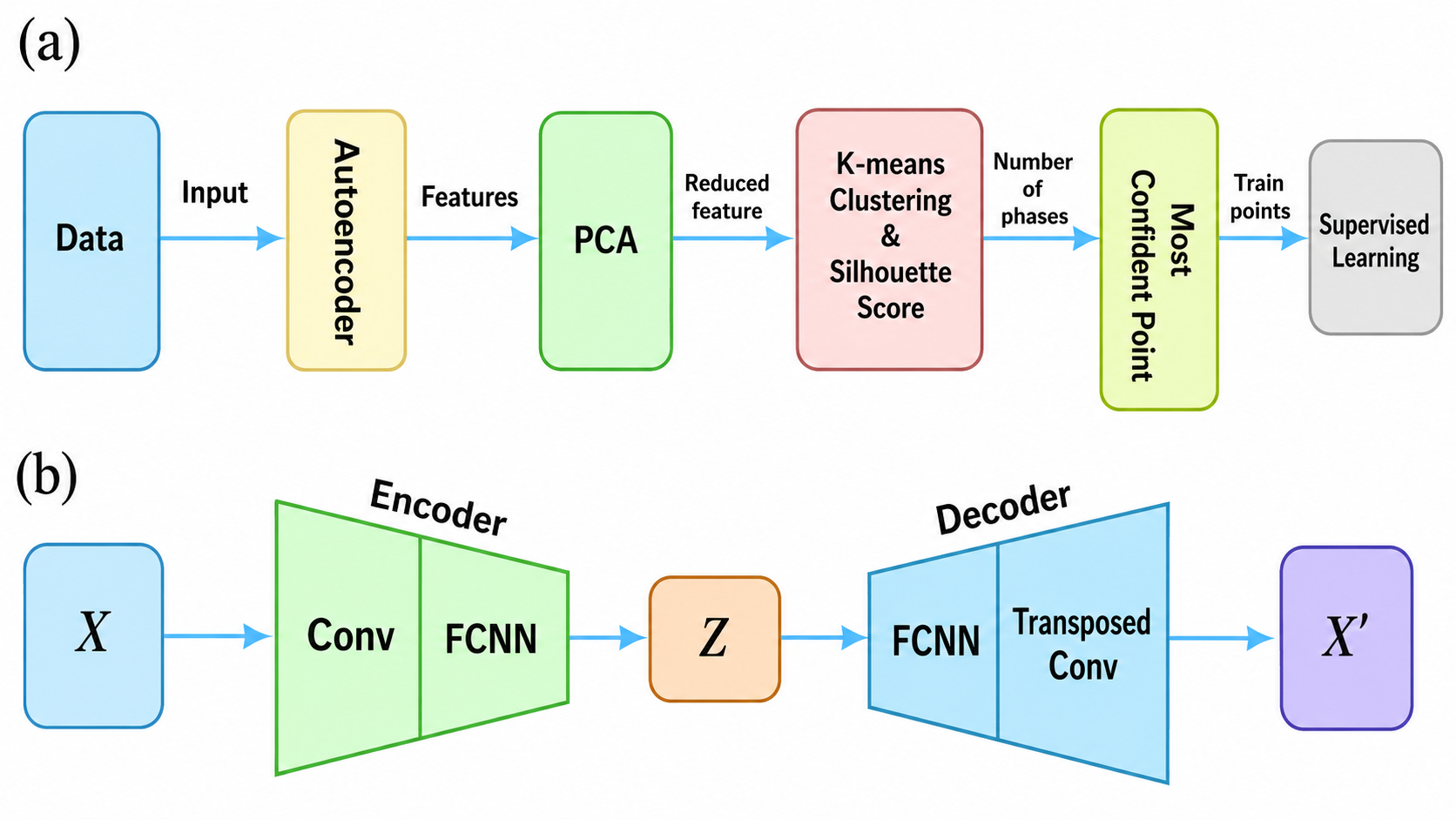}
    \caption{
    Schematic illustration of the machine-learning protocol.
    (a) Overall pipeline for extracting phase information from
    correlation matrices using autoencoder feature extraction, PCA,
    K-means clustering, silhouette analysis, and supervised refinement.
    (b) Autoencoder architecture used to compress the input correlation
    matrices into latent representations.
    }
    \label{fig:Pipeline}
  \end{center}
\end{figure}

The overall workflow~\cite{Ming21,Ming23} is summarized in Fig.~\ref{fig:Pipeline}.
Fig.~\ref{fig:Pipeline}(a) shows the complete machine-learning
pipeline, including autoencoder-based feature extraction, PCA
dimensional reduction, K-means clustering, silhouette analysis, and
supervised refinement. Fig.~\ref{fig:Pipeline}(b) illustrates the
autoencoder architecture used to compress the correlation matrices
into latent features.

Our protocol consists of several steps. For a review of the
algorithms involved, see Refs.~\cite{Goodfellow2016,Mehta2019}.
First, the correlation
matrices are fed into an autoencoder (AE)~\cite{Hinton2006}, which learns a compressed
latent representation of the input data in an unsupervised manner. The
encoder maps the original correlation matrix to a lower-dimensional
feature space, while the decoder reconstructs the input from this
compressed representation. After training, the latent variables
extracted by the encoder are taken as the effective features of the
system.

Principal component analysis (PCA)~\cite{Pearson1901,Jolliffe2002} is then applied to the latent
features in order to identify the dominant directions of variation and
to determine the effective dimensionality required to represent the
data. In practice, we keep the principal components that account for
99\% of the total variance, an empirical prescription adopted in our
previous work~\cite{Ming21}. The resulting reduced features are then
used for clustering.

To identify the phase structure without prior labels, we apply
K-means clustering~\cite{MacQueen1967,Lloyd1982} to the reduced feature space. Since the number of
clusters is not assumed beforehand, we evaluate the clustering quality
using silhouette analysis~\cite{Rousseeuw1987}. For a given partition of the
data into clusters, each data point $x$ is assigned a silhouette value
(SV)
\be
  s(x) = \frac{b(x)-a(x)}{\max\{a(x),b(x)\}},
\ee
where $a(x)$ is the mean distance between $x$ and all other points in
its own cluster, and $b(x)$ is the smallest mean distance from $x$ to
the points of any other cluster. The SV is bounded between $-1$ and
$1$ and measures how similar $x$ is to its own cluster (cohesion)
compared with the other clusters (separation). The mean of the SVs
over all data points, which we call the silhouette score, quantifies
the overall quality of the partition. For each trial number of
clusters $n$, the silhouette score is evaluated after K-means
clustering, and the optimal $n$ is chosen as the one that maximizes
it.

Once the cluster structure is obtained, the data points with the
largest SVs in each cluster are regarded as the most
representative and reliable samples of that cluster. These
high-confidence points can then be used to construct an effective
labeled set for a subsequent supervised refinement~\cite{Ming21}. Specifically,
for each cluster we (i) take the point with the largest SV as the
center of a high-confidence region; (ii) select two additional points,
one on each side of the center and as evenly spaced as possible,
within the region where the SV remains above $95\%$ of its maximum in
that cluster; and (iii) generate $1000$ additional correlation
matrices at equally spaced values of $\theta$ within a window of width
$0.01\pi$ around each of the three points, labeled by the cluster
index. The $95\%$ threshold is an empirical prescription that keeps the
training points well inside each phase, away from the transition
regions where the SV drops. The original $20001$ points serve as the
unlabeled test set on which the trained classifier predicts the phase
probabilities. Throughout this work, latent feature maps and
unsupervised phase assignments are shown for a representative model
whose behavior is shared by the majority of the $100$ independently
trained models, whereas the latent dimension, the silhouette scores,
and the predicted critical points are statistics over all $100$
runs. The unsupervised assignments serve only to select the number of
clusters and the high-confidence training points; the phase
boundaries reported below are determined by the subsequent supervised
refinement, whose statistics do not depend on the cluster labels.
This final step is
used only to improve the precision of the phase-boundary
determination; the initial identification of the phase structure
itself remains unsupervised.

\subsection{Data preparation}
\label{sec:data}
Using iDMRG~\cite{McCulloch2008,Schollwock2011}, we compute the four
correlation matrices $C_{\mathrm{SS}}$, $C_{\mathrm{DM}}$,
$C_{\mathrm{TM}}$, and $C_{\mathrm{SU(3)}}$ at $20001$ uniformly
spaced values of $\theta$ in $[0,2\pi)$, with a block size $l_s=10$
and a truncation bond dimension $m_{\mathrm{tr}}=100$. The iDMRG
calculations employ an $SU(2)$-symmetric matrix product state with a
unit cell of $10$ sites, and the correlation matrices are evaluated
within one unit cell, so that $l_s$ coincides with the unit-cell
length. In the ferromagnetic region, enforcing $SU(2)$ symmetry
constrains the finite-bond-dimension MPS to a symmetry-preserving
approximation. Such states can closely reproduce the local
ferromagnetic correlations while exhibiting weakly damped,
long-wavelength oscillations at larger separations. These oscillations
are features of the constrained approximation and should not be
interpreted as physical modulation of the ferromagnetic ground
state. The relevant numerical accuracy for the present analysis is
therefore that of the correlation matrices over the finite input
window ($l_s=10$). The weak $\theta$ dependence of the oscillation
wavelength across the ferromagnetic phase is the origin of the small
scatter of the SVs in that phase noted below. The calculations are iterated until the relative change of
the energy per site between successive sweeps falls below $10^{-7}$;
the adequacy of $m_{\mathrm{tr}}=100$ was verified on a coarser grid
of $200$ values of $\theta$ before the full scan. The correlations are
unconnected, $\langle O_iO_j\rangle$, and include the diagonal
elements $i=j$; for the two- and three-site operators $D_i$ and $T_i$
the operator is required to lie entirely within the block, giving
$9\times9$ and $8\times8$ matrices that are zero-padded to
$10\times10$. In the dimerized phase, iDMRG converges to either of
the two translation-related ground states, and both dimer patterns
are present in the data. Each correlation channel is rescaled to
$[0,1]$ by min-max normalization over the whole data set before
training.
To examine the role of each correlation function in distinguishing the
different phases, we begin with the full set of four correlation
matrices and then systematically remove one, two, and three
correlations. The results of these correlation-ablation analyses are
discussed in the following subsections.

\subsection{Unsupervised results}
\label{sec:results}

\subsubsection{4 correlations }
\begin{figure}[th]
  \begin{center}
    \includegraphics[width=9cm]{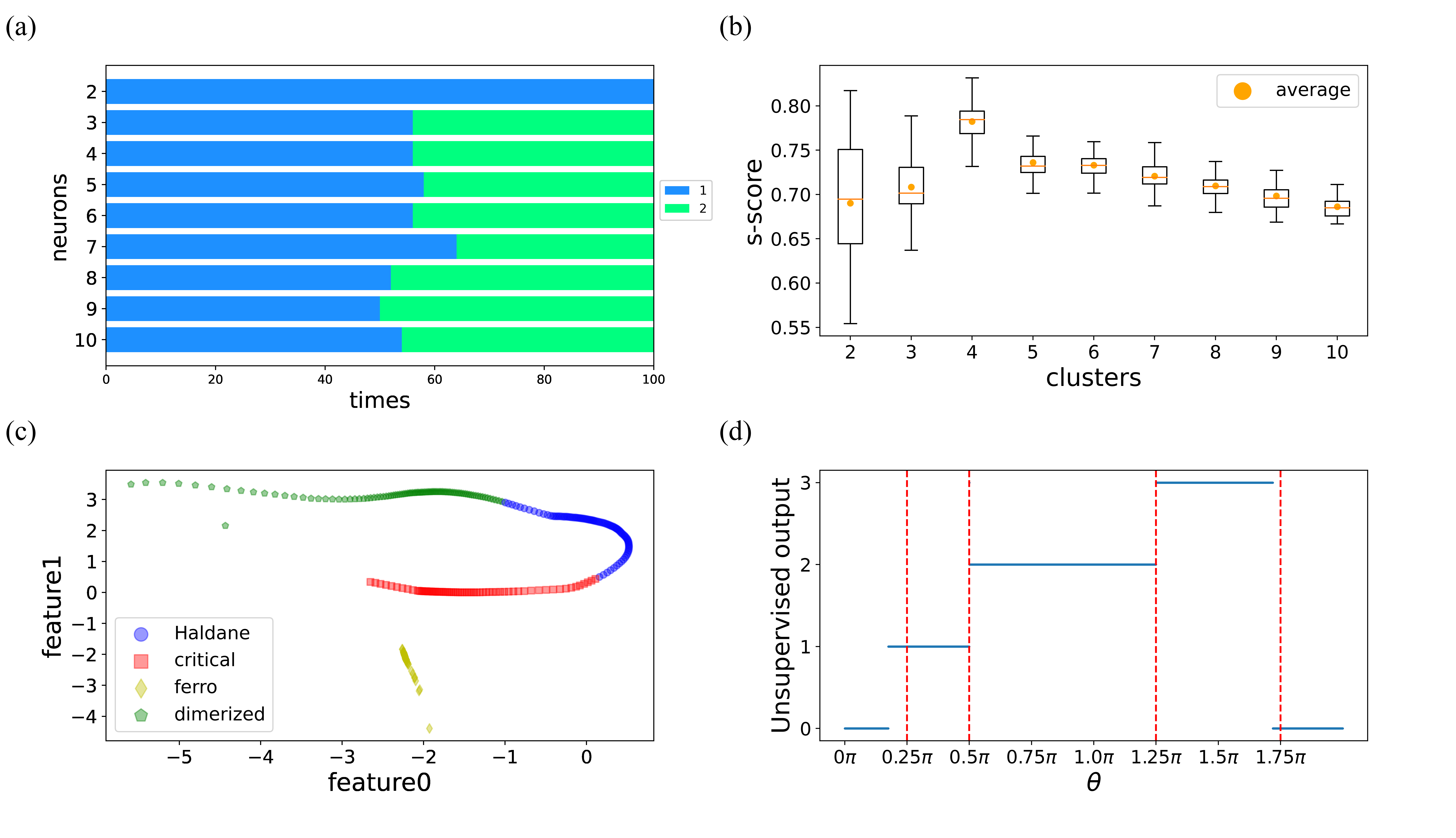}
    \caption{AE results using all four correlation functions,
$C_{\mathrm{SS}}(r)$, $C_{\mathrm{DM}}(r)$,
$C_{\mathrm{TM}}(r)$, and $C_{\mathrm{SU(3)}}(r)$, as input data for
the BLBQ model.
(a) Discrete distribution of the necessary number of latent neurons
$d_z$ for a given number of neurons $n_{\mathrm{mid}}$ in the
middlemost layer, with $n_{\mathrm{mid}}=2,\ldots,10$ along the
vertical axis. The statistics are obtained from $100$ independently
trained AEs. The length of each color bar is proportional to the
number of times that a given $d_z$ occurs among the $100$ models, and
different colors denote different values of $d_z$.
(b) Box plot of the silhouette scores as a function of the number of
clusters $n$ obtained from K-means clustering.
(c) Latent representations in the basis of the two principal
components. Each color denotes the corresponding
cluster, or phase. The discontinuous transitions into the FM phase are reflected by
disconnected clusters, whereas the continuous transitions are
represented by smoothly connected clusters.
(d) Phase assignments obtained from the unsupervised-learning
pipeline for a representative model (see Sec.~\ref{sec:protocol}).
}
    
    \label{fig:AEResults_FMDMTMSU3}
  \end{center}
\end{figure}

We first train the autoencoder to encode the correlation-function
data into latent representations. The minimal dimension $d_z$ of the
latent space is determined as follows. For each fixed number of hidden
neurons $n_{\mathrm{mid}}$ in the middlemost layer, with
$n_{\mathrm{mid}}=2,\ldots,10$, we independently train $100$ AEs using
the same model architecture but different random initializations of the
network weights.
As shown in Fig.~\ref{fig:AEResults_FMDMTMSU3}(a), we then compute
the discrete distribution of $d_z$, defined as the number of principal
components required to retain at least $99\%$ of the variance of the
encoded representation vectors. For the input data constructed from
all four correlation functions, the analysis suggests $d_z=2$.  

After $d_z$ is determined, we set the number of neurons in the
middlemost layer to $n_{\mathrm{mid}}=d_z$ and use the corresponding
latent representations extracted from the trained AEs for silhouette
analysis. K-means clustering is then performed for different numbers
of clusters $n$, and the optimal value of $n$ is estimated from the
silhouette score. As shown in Fig.~\ref{fig:AEResults_FMDMTMSU3}(b),
the mean silhouette score reaches its maximum at $n=4$, indicating
that the correlation data naturally separate into four clusters. This
result is consistent with the existence of four phases in the BLBQ model.

The two-dimensional latent representations are shown in
Fig.~\ref{fig:AEResults_FMDMTMSU3}(c) in the basis of the two
principal components. For the input data
constructed from all four correlation functions, the feature map exhibits
four distinguishable clusters. The yellow-green points,
corresponding to the ferromagnetic (FM) phase, are discontinuously
separated from the other clusters. This discontinuity is consistent
with the discontinuous transitions at $\theta_c=\pi/2$ and
$\theta_c=5\pi/4$, the latter being direct but not conventionally
first order (Sec.~\ref{sec:BLBQ}). In contrast, the green dimerized (DM), blue
Haldane (HP), and red trimerized (TM) clusters are connected
continuously, reflecting the continuous transitions at
$\theta_c=-\pi/4$ and $\theta_c=\pi/4$.

\begin{figure}[th]
  \begin{center}    \includegraphics[width=9cm]{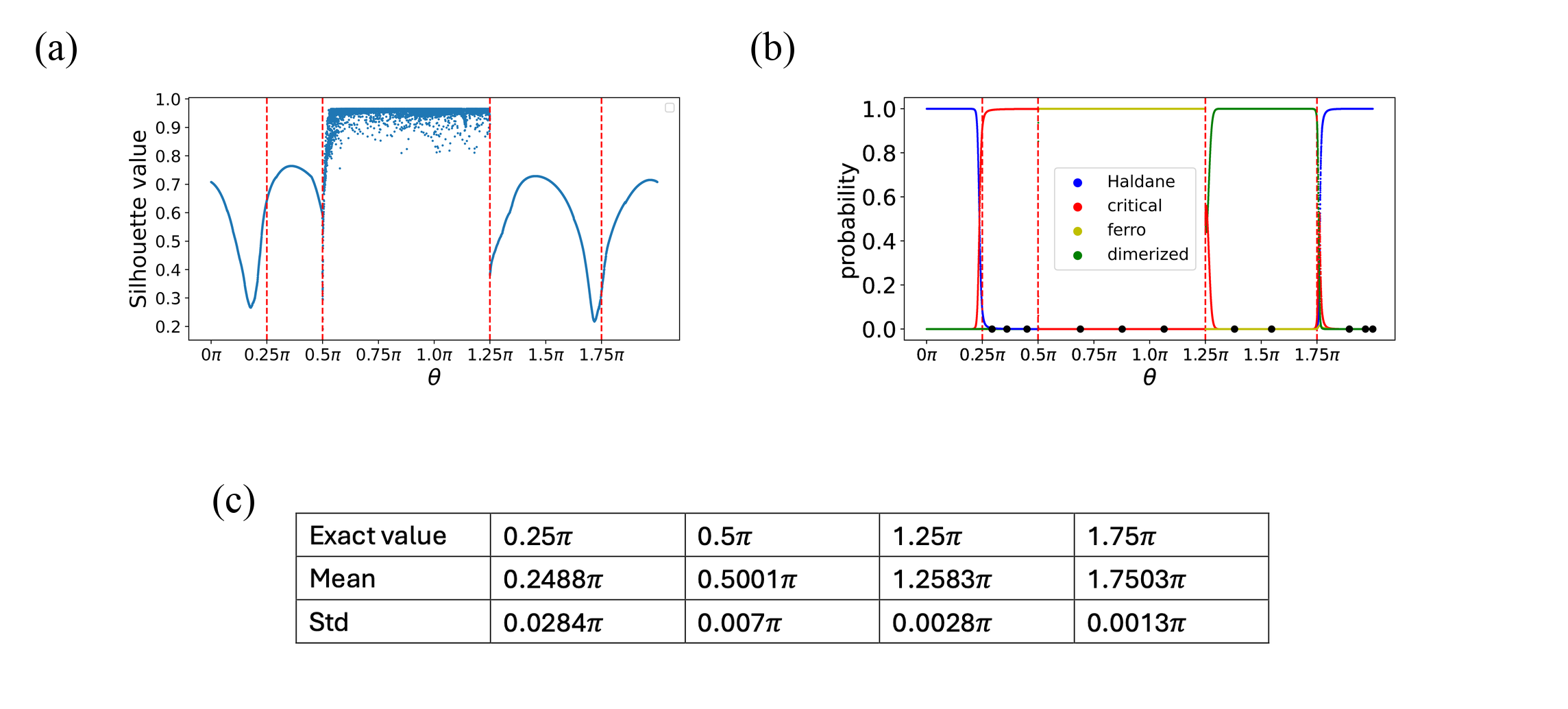}
    \caption{
Results of the supervised-learning refinement using the
high-confidence points selected from the silhouette analysis.
(a) Silhouette values (SVs) as a function of $\theta$. The red
dashed lines indicate the exact phase-transition points. The
$2$ continuous minima of
the SV are located close to, but not exactly at, the continuous phase
transitions at $\theta=\pi/4$ and $\theta=7\pi/4$, while the SV
remains nearly constant within the ferromagnetic phase.
(b) Probabilities predicted by the supervised neural network for the
four phases. The phase-transition points are determined from the
crossings of the probability curves. The red dashed lines denote the
exact critical points. Black dots mark the training points, at
$\theta/\pi=0.2931$, $0.3601$, $0.4497$ (critical trimerized);
$0.6875$, $0.8750$, $1.0625$ (ferromagnetic); $1.3802$, $1.4545$,
$1.5467$ (dimerized); and $1.8948$, $1.9675$, $1.9989$ (Haldane).  
(c) Exact critical points, mean predicted values, and standard
deviations over $100$ independent ML runs. The predicted
values agree with the exact ones to within $0.01\pi$; the
ferromagnetic-dimerized boundary is discussed in the text.
}
    \label{fig:SCSV_FMDMTMSU3} 
  \end{center}
\end{figure}

Fig.~\ref{fig:AEResults_FMDMTMSU3}(d) shows the phase assignments
obtained from K-means clustering with the selected feature dimension
and the optimal number of clusters, shown for a representative
model whose behavior is shared by the majority of the $100$ runs. The
unsupervised results already reproduce the overall phase structure
with good accuracy. To further refine the precision of the
phase-boundary determination, we use the data points with the largest
SV in each cluster as high-confidence samples for
supervised learning. The refined results are shown in
Fig.~\ref{fig:SCSV_FMDMTMSU3}.

The SVs are shown in
Fig.~\ref{fig:SCSV_FMDMTMSU3}(a). In the ferromagnetic phase, the SV
forms an almost flat profile with small random fluctuations, which
we attribute to the weak $\theta$ dependence of the long-wavelength
oscillation of the $SU(2)$-symmetric MPS approximation
(Sec.~\ref{sec:data}). Similar
behavior also appears in other ferromagnetic phases, such as the FM
phase of the XXZ model discussed in our previous work
\cite{Ming23}. The two continuous minima of the SV are located close to, but not exactly at, the
critical points. Since the SV minima mark the cluster boundaries
found by K-means from distances in feature space alone, and the
correlation matrices evolve smoothly across a continuous transition,
these boundaries need not coincide exactly with the critical
points. The displacement is thus expected for the continuous
transitions at $\theta=\pi/4$ and $7\pi/4$, whereas the discontinuous
transitions into the ferromagnetic phase appear as sharp jumps of the
SV. The larger displacement at $\theta=\pi/4$ is consistent with the
Berezinskii-Kosterlitz-Thouless character of the ULS transition, near
which the correlation matrices change particularly slowly.

The training points, selected as described in
Sec.~\ref{sec:protocol}, are marked by black dots in
Fig.~\ref{fig:SCSV_FMDMTMSU3}(b).

The results of the supervised learning are shown in
Fig.~\ref{fig:SCSV_FMDMTMSU3}(b) and (c). The $y$-axis of
Fig.~\ref{fig:SCSV_FMDMTMSU3}(b) represents the probability that a
given point belongs to a particular phase. The phase-transition points
are identified by the crossings of the corresponding probability
curves.

The predicted critical points and their standard deviations, obtained
from $100$ independent ML runs, are listed in
Fig.~\ref{fig:SCSV_FMDMTMSU3}(c). The predicted values agree with the
exact critical points to within $0.01\pi$. For three of the four
boundaries the deviation is smaller than the standard deviation over
the $100$ independent training runs, which measures the stability of
the training at fixed input data rather than a physical uncertainty;
the comparatively large spread at $\theta=\pi/4$ reflects the slow
variation of the correlations near the BKT transition. The exception
is the
ferromagnetic-dimerized boundary, predicted at
$1.2583\pi\pm0.0028\pi$: its deviation from $5\pi/4$ is about three
standard deviations and therefore a systematic bias. This is expected
from the nature of this transition discussed in Sec.~\ref{sec:BLBQ}:
the characteristic length scale grows very large on the dimerized
side, so that with a finite block size $l_s$ and bond dimension a
narrow portion of the dimerized phase retains correlation patterns
resembling the ferromagnetic endpoint. Consistently, the predicted
boundary lies on the dimerized side of $5\pi/4$, and it should be read
as a finite-resolution estimate rather than a measurement of the
thermodynamic boundary.

\subsubsection{3 Correlations}

\begin{figure}[th]
  \begin{center}
    \includegraphics[width=9cm]{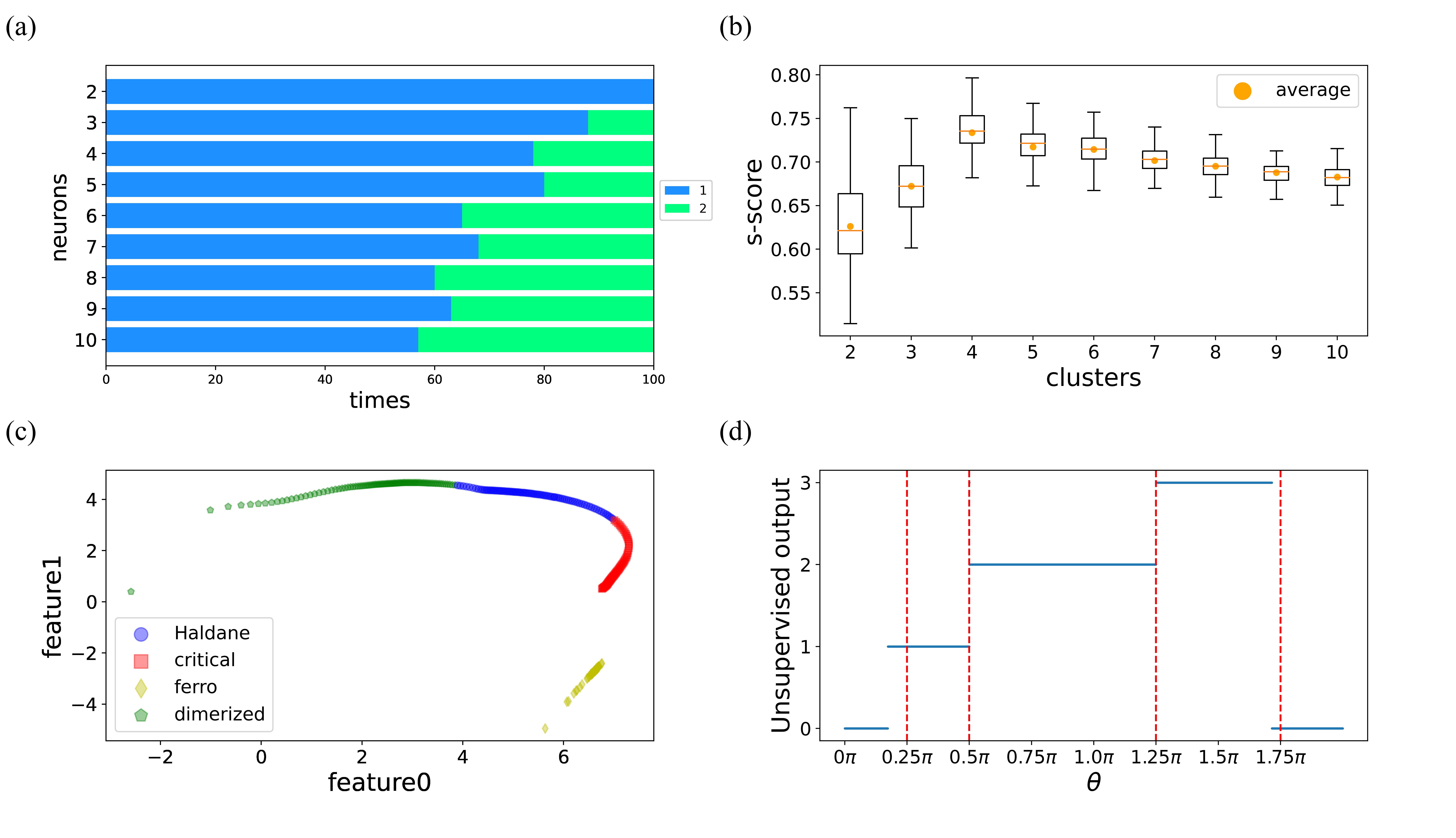}
    \caption{AE results using three correlation functions,
$C_{\mathrm{SS}}(r)$, $C_{\mathrm{DM}}(r)$, and
$C_{\mathrm{SU(3)}}(r)$, with the trimer-trimer correlation
$C_{\mathrm{TM}}(r)$ removed. Panels (a), (b), (c), and (d) are the
same as those in Fig.~\ref{fig:AEResults_FMDMTMSU3}.
}
 \label{fig:AEResults_FMDMSU3}
  \end{center}
\end{figure}

In order to investigate the role of each correlation function in
determining the critical points, we systematically remove one
correlation function at a time and use the remaining three
correlations as inputs to the machine-learning pipeline. Since the
trimer-trimer correlations and the $SU(3)$ correlations both encode
the period-$3$ structure of the critical (trimerized) phase, some
overlap of information is expected between them. We therefore begin by
removing the trimer-trimer correlations and use
$C_{\mathrm{SS}}(r)$, $C_{\mathrm{DM}}(r)$, and
$C_{\mathrm{SU(3)}}(r)$ as the input data. The unsupervised results
are shown in Fig.~\ref{fig:AEResults_FMDMSU3}.

As shown in Fig.~\ref{fig:AEResults_FMDMSU3}(a) and (b), the latent
dimension is again $d_z=2$ and the silhouette analysis again selects
$n=4$, so that the correct number of phases is recovered when the
trimer-trimer correlation is removed.

Fig.~\ref{fig:AEResults_FMDMSU3}(c) and (d) show the latent feature
map and the corresponding unsupervised phase assignments,
respectively. The results are qualitatively similar to those obtained
from the full set of four correlation functions.

\begin{figure}[th]
  \begin{center}
    \includegraphics[width=9cm]{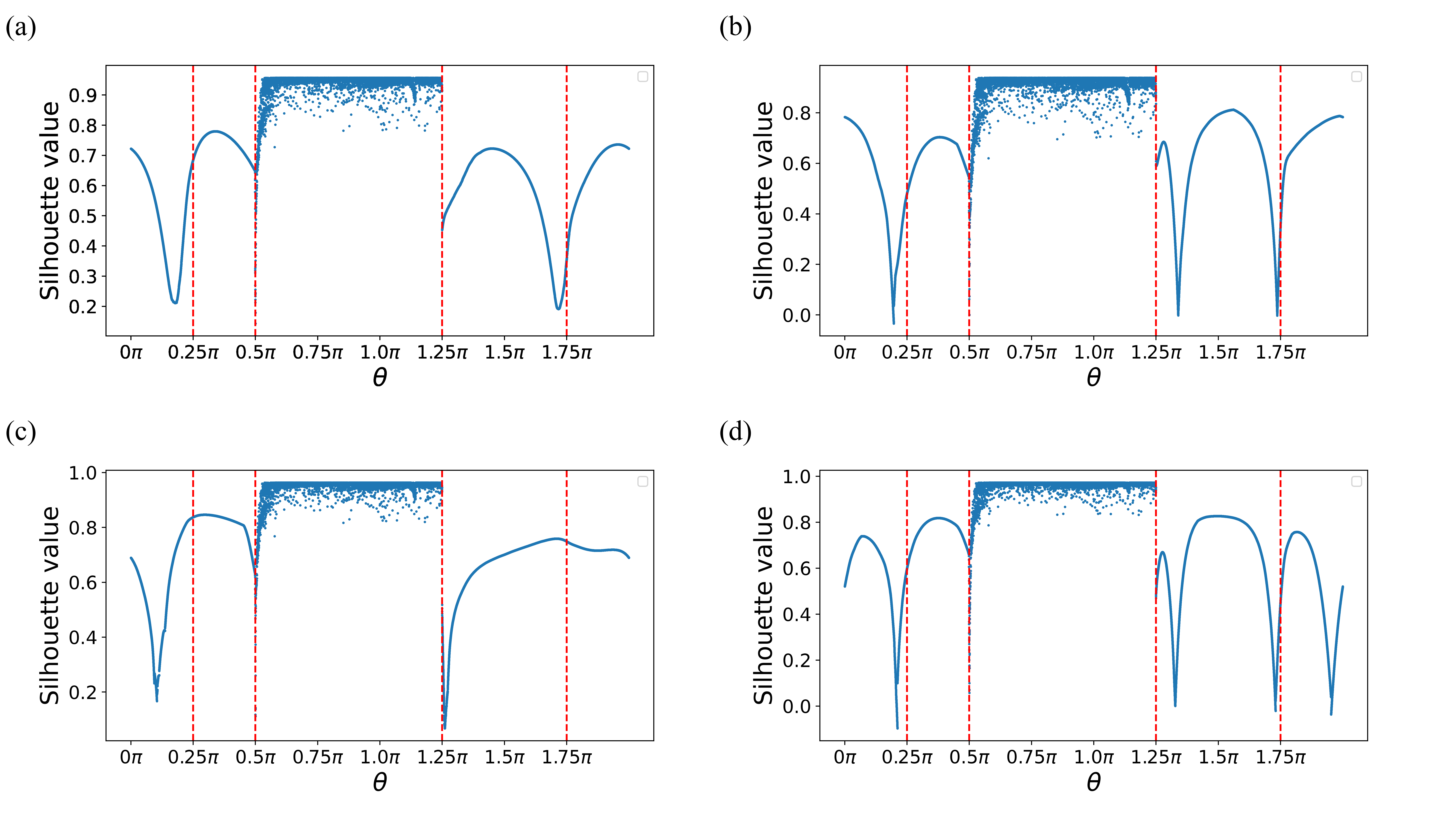}
    \caption{
Silhouette values obtained from the three-correlation inputs.
(a) $C_{\mathrm{TM}}(r)$ removed.
(b) $C_{\mathrm{SU(3)}}(r)$ removed.
(c) $C_{\mathrm{DM}}(r)$ removed.
(d) $C_{\mathrm{SS}}(r)$ removed.
The red dashed lines indicate the exact phase-transition points. The
local minima of the silhouette values provide useful information about
the phase boundaries and possible phase grouping.}
    \label{fig:SV_CF3}
  \end{center}
\end{figure}

The SVs obtained with $C_{\mathrm{TM}}(r)$ removed are
shown in Fig.~\ref{fig:SV_CF3}(a). The ferromagnetic phase still
exhibits an almost flat SV profile. The minima of the SVs also remain
located near the continuous phase-transition points. However, the
distances between the minima and the exact critical points are larger
than those obtained using all four correlation functions. This
suggests that the subsequent supervised learning is expected to yield
less precise estimates of the critical points.

\begin{figure}[th]
  \begin{center}
    \includegraphics[width=9cm]{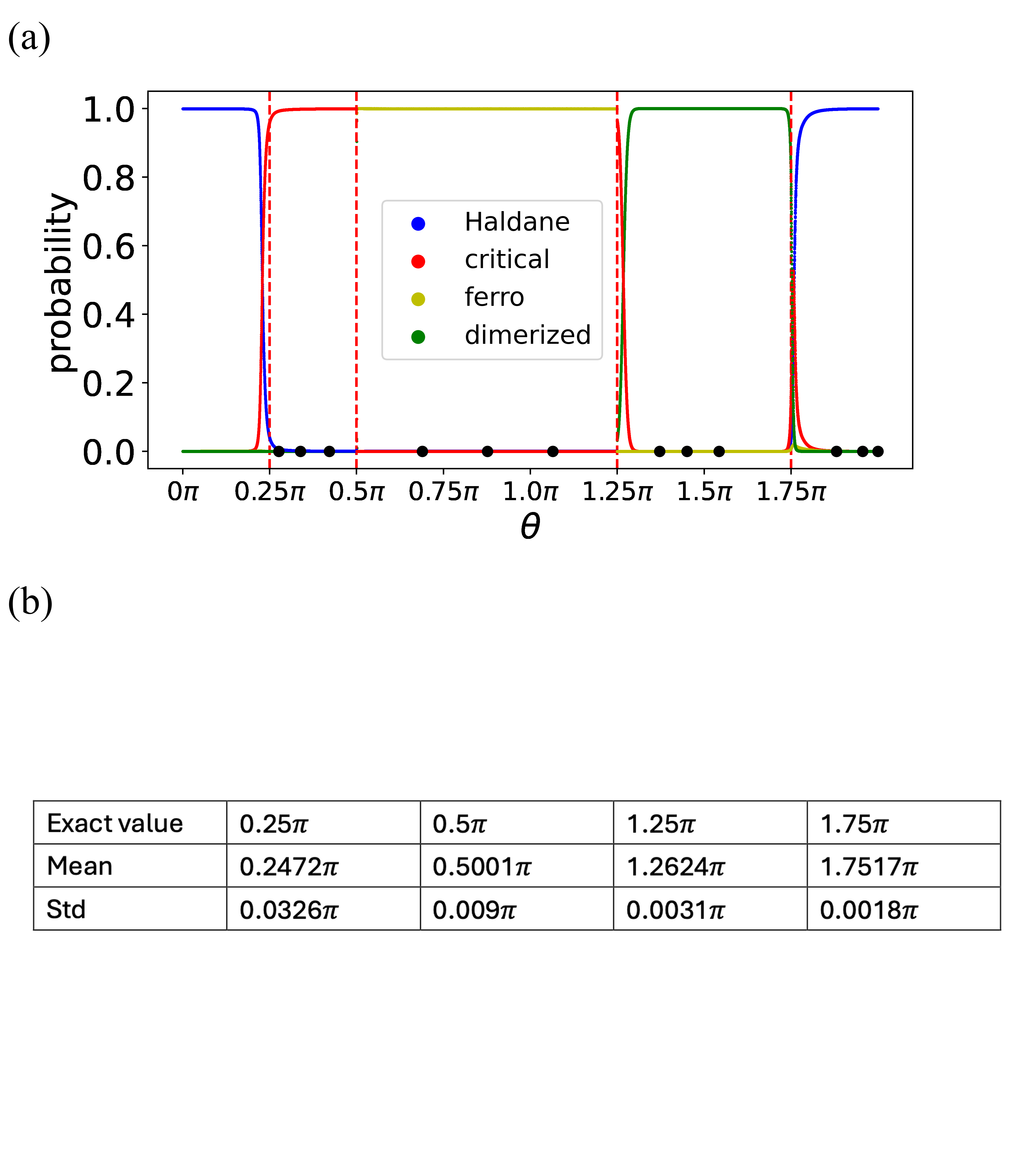}
    \caption{
   (a) Supervised-learning refinement using three correlation functions,
    with $C_{\mathrm{TM}}(r)$ removed. The black dots indicate the
    high-confidence training points selected from the silhouette
    values, at $\theta/\pi=0.2759$, $0.3863$, $0.4214$ (critical
    trimerized); $0.6875$, $0.8750$, $1.0625$ (ferromagnetic); $1.3720$,
    $1.4507$, $1.5432$ (dimerized); and $1.8809$, $1.9556$, $1.9984$
    (Haldane). (b) Predicted phase-transition points.
    }
    \label{fig:Supervised_FMDMSU3}
  \end{center}
\end{figure}

Fig.~\ref{fig:Supervised_FMDMSU3} shows the critical points obtained
from the supervised-learning refinement for the three-correlation
input data, with $C_{\mathrm{TM}}(r)$ removed. The black dots indicate
the training points, selected as described in Sec.~\ref{sec:protocol}.

The predicted critical points and their standard deviations over
$100$ independent ML runs are listed in
Fig.~\ref{fig:Supervised_FMDMSU3}(b) and still agree well with the
exact values. Compared with the four-correlation results, however,
both the deviations from the exact values and the standard deviations
are slightly larger, suggesting that the trimer-trimer correlation
carries some additional, though not essential, information for
locating the phase boundaries.

We next examine the remaining three-correlation combinations.
Fig.~\ref{fig:SV_CF3}(b), (c), and (d) show the corresponding SVs
with $C_{\mathrm{SU(3)}}(r)$, $C_{\mathrm{DM}}(r)$, and
$C_{\mathrm{SS}}(r)$ removed, respectively. When
$C_{\mathrm{SU(3)}}(r)$ is removed, a new local minimum appears
between the FM and DM phases, suggesting that an additional phase may
be incorrectly identified between these two phases.

When $C_{\mathrm{DM}}(r)$ is removed, the local minimum between the
Haldane and DM phases disappears, suggesting that the Haldane phase is
expected to be merged with neighboring phases in the subsequent
clustering. In contrast, when $C_{\mathrm{SS}}(r)$ is removed, one
additional local minimum appears inside the DM phase, close to its
boundary with the FM phase, and another inside the Haldane phase, so
that the SV curve indicates six regions. This makes the phase
identification considerably more complicated.

\begin{figure}[th]
  \begin{center}
    \includegraphics[width=9cm]{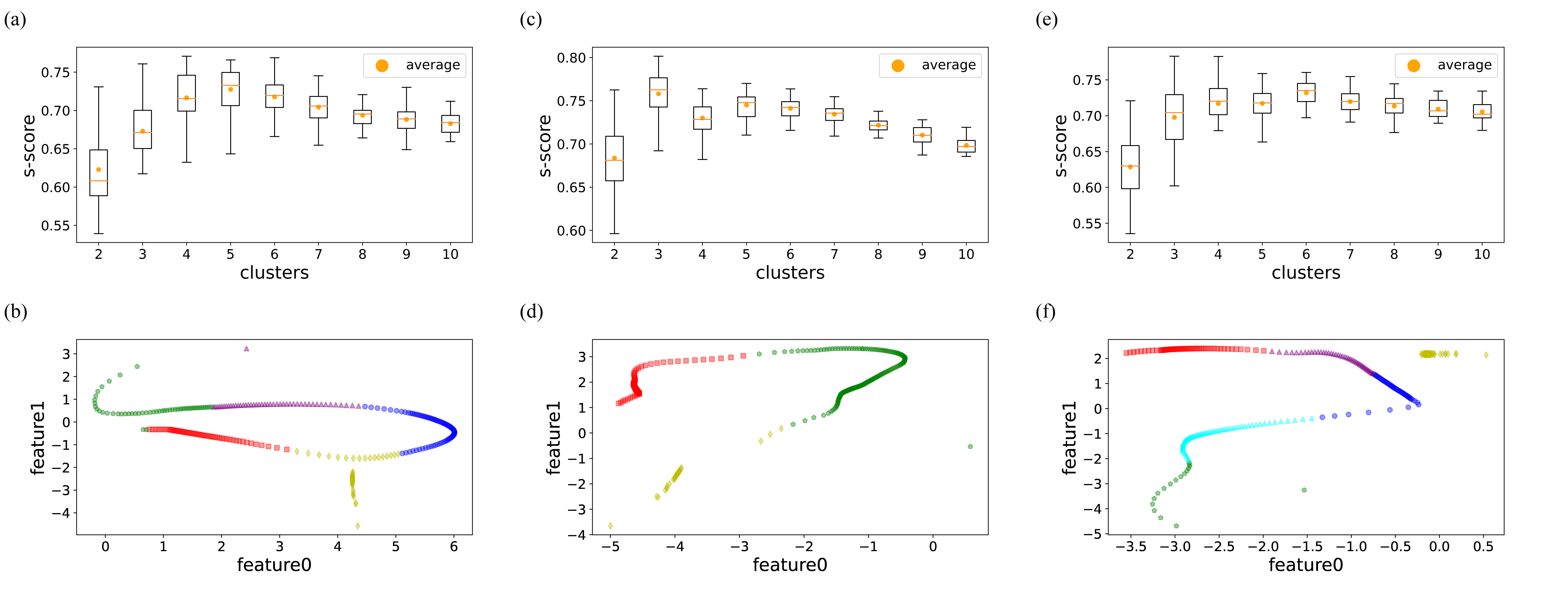}
    \caption{
    Silhouette scores and latent feature maps for the
    three-correlation inputs with incorrect phase grouping. Panels
    (a) and (b) correspond to $C_{\mathrm{SU(3)}}(r)$ removed,
    panels (c) and (d) correspond to $C_{\mathrm{DM}}(r)$ removed,
    and panels (e) and (f) correspond to
    $C_{\mathrm{SS}}(r)$ removed.
    }
    \label{fig:US_FM_3CF_notCorrect}
  \end{center}
\end{figure}

Fig.~\ref{fig:US_FM_3CF_notCorrect} shows the silhouette scores and
latent feature maps obtained from the unsupervised-learning analysis.
Fig.~\ref{fig:US_FM_3CF_notCorrect}(a) and (b) correspond to the case
with $C_{\mathrm{SU(3)}}(r)$ removed.
Fig.~\ref{fig:US_FM_3CF_notCorrect}(c) and (d) correspond to the case
with $C_{\mathrm{DM}}(r)$ removed, while
Fig.~\ref{fig:US_FM_3CF_notCorrect}(e) and (f) correspond to the case
with $C_{\mathrm{SS}}(r)$ removed.

In all three cases, the silhouette scores no longer suggest the
correct number of clusters. When $C_{\mathrm{SU(3)}}(r)$,
$C_{\mathrm{DM}}(r)$, and $C_{\mathrm{SS}}(r)$ are removed,
respectively, the largest average silhouette scores occur at
$n=5$, $n=3$, and $n=6$, respectively. In other words, the
machine-learning pipeline incorrectly suggests that the system should
be divided into $5$, $3$, and $6$ phases, respectively.

As discussed above, the results $n=5$, $n=3$, and $n=6$ for the cases
with $C_{\mathrm{SU(3)}}(r)$, $C_{\mathrm{DM}}(r)$, and
$C_{\mathrm{SS}}(r)$ removed, respectively, are consistent with the
structures indicated by the SVs. In the last case, however, the
silhouette scores for $n=4$ to $7$ are very close to one another,
indicating strong competition among different clustering
configurations; the selected $n=6$ is therefore less robust than in
the other cases.

A closer inspection of the latent feature maps helps explain how these
incorrect cluster numbers arise. When $C_{\mathrm{SU(3)}}(r)$ is
removed, an additional cluster appears between the ferromagnetic and
dimerized phases. When $C_{\mathrm{DM}}(r)$ is removed, the Haldane
phase is no longer identified as a separate cluster; instead, part of
it is grouped with the critical phase and another part with the
dimerized phase. As a result, the Haldane phase effectively
disappears from the unsupervised clustering. Finally, when
$C_{\mathrm{SS}}(r)$ is removed, two additional clusters emerge: one
between the ferromagnetic and dimerized phases, and another between
the dimerized and Haldane phases. This observation is consistent with
the strong competition among different clustering configurations
discussed above.

\begin{figure}[th]
  \begin{center}
    \includegraphics[width=9cm]{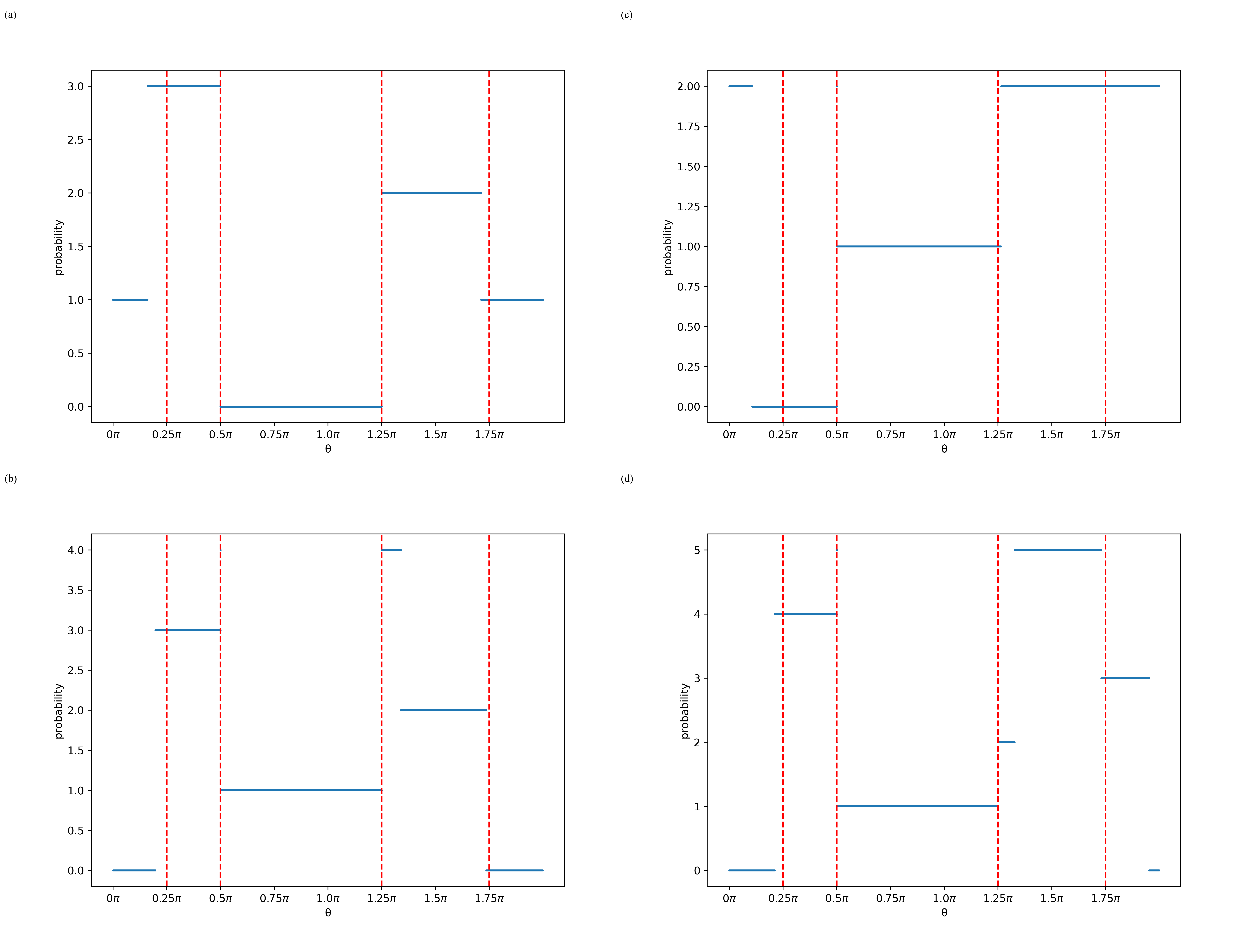}
    \caption{
    Unsupervised phase assignments obtained from the latent
    representations. Panels (a), (b), (c), and (d) correspond to
    $C_{\mathrm{TM}}(r)$, $C_{\mathrm{SU(3)}}(r)$,
    $C_{\mathrm{DM}}(r)$, and $C_{\mathrm{SS}}(r)$ removed,
    respectively. The red dashed lines indicate the exact
    phase-transition points, and the cluster indices on the vertical
    axis are arbitrary.
    }
    \label{fig:USCP_FM_3CF_notCorrect}
  \end{center}
\end{figure}

This behavior becomes clearer when we examine the unsupervised phase
assignments shown in Fig.~\ref{fig:USCP_FM_3CF_notCorrect}. The four
cases are shown in Fig.~\ref{fig:USCP_FM_3CF_notCorrect}(a)-(d),
corresponding to the removal of $C_{\mathrm{TM}}(r)$,
$C_{\mathrm{SU(3)}}(r)$, $C_{\mathrm{DM}}(r)$, and
$C_{\mathrm{SS}}(r)$, respectively.
The output labels assigned by the unsupervised algorithm are arbitrary
and therefore not directly comparable among different cases. However,
the resulting phase structures still reveal which phases become
redundant or are split into additional clusters.

When $C_{\mathrm{TM}}(r)$ is removed, the unsupervised result still
correctly reproduces the four-phase structure. When
$C_{\mathrm{SU(3)}}(r)$ is removed, an additional phase appears
between the FM and DM phases and is identified as the fourth output
cluster. When $C_{\mathrm{DM}}(r)$ is removed, the Haldane phase
disappears. These three cases are consistent with the behavior
suggested by the SVs in
Fig.~\ref{fig:SV_CF3}(a), (b), and (c), respectively.

When $C_{\mathrm{SS}}(r)$ is removed, two additional phases appear:
one between the FM and DM phases and another between the DM and
Haldane phases, identified as the second and third output clusters,
respectively.

To summarize the results for the three-correlation inputs, removing
the trimer-trimer correlation has only a minor effect on the
predicted critical points. The resulting phase boundaries are only
slightly less accurate than those obtained using all four correlations.
In contrast, when the trimer-trimer correlation is retained but one of
the other correlations is removed, the predicted phase grouping no
longer agrees with the exact four-phase structure. Depending on which
correlation is removed, the machine-learning pipeline predicts either
one or two additional phases, or fails to identify one of the existing
phases.

At this stage, however, the three-correlation results are still not
sufficient to determine the precise role of each individual
correlation function. In the next subsection, we therefore consider
two-correlation inputs, which provide more detailed information about
the strengths and limitations of different correlation combinations.
The role of each single correlation function will become clearest only
when one-correlation inputs are examined.

\subsubsection{2 Correlations}

 We now consider the cases in which only two correlation functions are
used as inputs to the machine-learning pipeline. As before, we first
examine the best-performing combinations and then proceed to those
with progressively deteriorating performance.

\begin{figure}[th]
  \begin{center}
    \includegraphics[width=9cm]{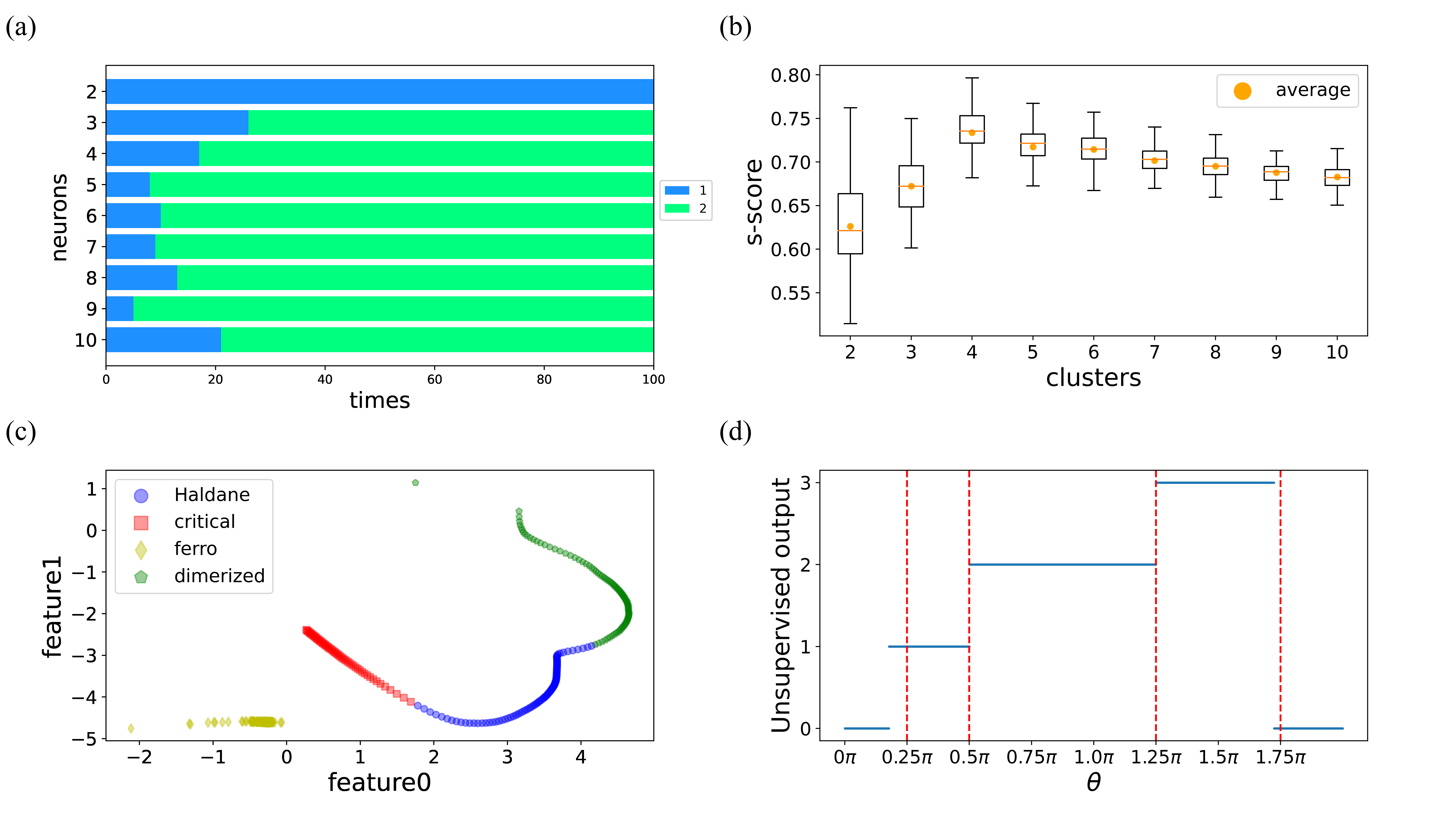}
    \caption{AE results using two correlation functions,
$C_{\mathrm{SS}}(r)$ and $C_{\mathrm{DM}}(r)$. Panels (a), (b), (c), and (d) are the
same as those in Fig.~\ref{fig:AEResults_FMDMTMSU3}.
}
 \label{fig:AEResults_FMDM}
  \end{center}
\end{figure}

\begin{figure}[th]
  \begin{center}
    \includegraphics[width=9cm]{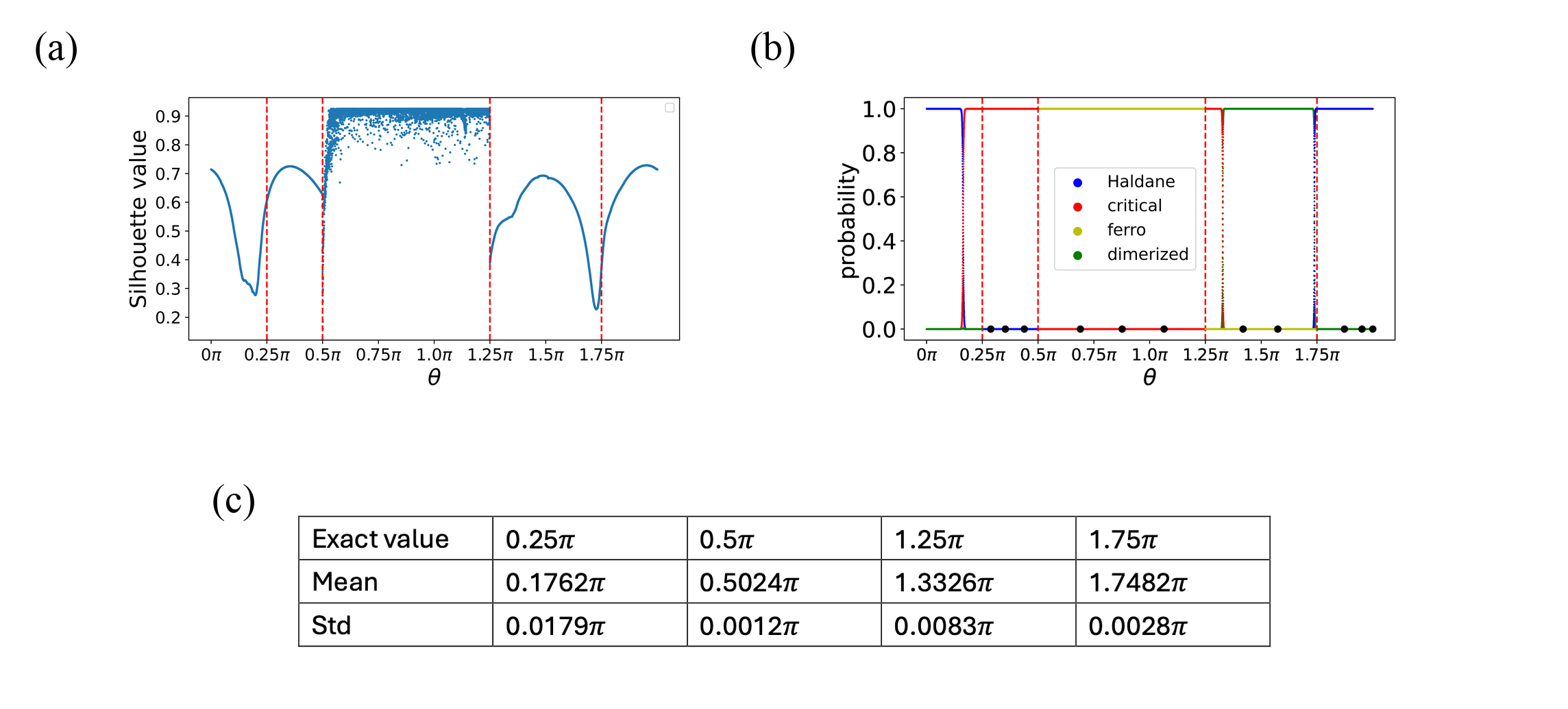}
    \caption{
Results obtained using two correlation functions,
$C_{\mathrm{SS}}(r)$ and $C_{\mathrm{DM}}(r)$.
(a) Silhouette values. (b) Supervised-learning refinement based
on the high-confidence training points selected from the silhouette
values. The black dots indicate the selected training points, at
$\theta/\pi=0.2876$, $0.3543$, $0.4382$ (critical trimerized);
$0.6875$, $0.8750$, $1.0625$ (ferromagnetic); $1.4189$, $1.4879$,
$1.5744$ (dimerized); and $1.8726$, $1.9516$, $1.9789$ (Haldane).
(c) Predicted phase-transition points.}
    \label{fig:SCSV_FMDM}
  \end{center}
\end{figure}

The best results among the two-correlation inputs are
obtained from the combination of $C_{\mathrm{SS}}(r)$ and $C_{\mathrm{DM}}(r)$.
The unsupervised results are shown in Fig.~\ref{fig:AEResults_FMDM}. As
shown in Fig.~\ref{fig:AEResults_FMDM}(a) and (b), the latent
dimension is again $d_z=2$ and the silhouette analysis again selects
$n=4$, so that the correct number of phases is recovered from these
two correlations alone.

Fig.~\ref{fig:AEResults_FMDM}(c) and (d) show the latent feature map
and the corresponding unsupervised phase assignments, respectively.
The results are qualitatively similar to those obtained using all four
correlations, as well as those obtained from the three-correlation
input with $C_{\mathrm{TM}}(r)$ removed.

Fig.~\ref{fig:SCSV_FMDM}(a) shows the SVs obtained
using $C_{\mathrm{SS}}(r)$ and $C_{\mathrm{DM}}(r)$ as the input data.
The training points, selected as described in
Sec.~\ref{sec:protocol}, are marked by black dots in
Fig.~\ref{fig:SCSV_FMDM}(b).

The results of the supervised learning are shown in
Fig.~\ref{fig:SCSV_FMDM}(b) and (c), where the predicted critical
points and their standard deviations over $100$ independent ML runs
are listed in Fig.~\ref{fig:SCSV_FMDM}(c).

The predicted critical point between the critical and ferromagnetic
phases ($0.5\pi$), as well as that between the dimerized and Haldane phases ($1.75\pi$),
remains in excellent agreement with the exact values. However, the
predicted critical points between the Haldane  and critical trimerized
phases ($0.25 \pi$) 
and between the ferromagnetic and dimerized phases ($1.25 \pi$) are significantly
less accurate than those obtained using four correlations or three
correlations with $C_{\mathrm{TM}}(r)$ removed.

Signs of this reduced accuracy can already be observed in the SV
curve shown in Fig.~\ref{fig:SCSV_FMDM}(a). Around the
less-accurately predicted critical points, the SV curve exhibits
noticeable non-smooth behavior. Consistently, the silhouette scores for
$n=4$, $n=5$, and $n=6$ in Fig.~\ref{fig:AEResults_FMDM}(b) are very
close to one another. This indicates that although the machine
ultimately identifies four clusters as the optimal solution, the
underlying clustering structure is nearly degenerate. If the system is
partitioned into five or six clusters instead, the non-smooth regions
are further separated into additional phases. This behavior is similar
to that observed in the three-correlation results with
$C_{\mathrm{SS}}(r)$ or $C_{\mathrm{DM}}(r)$ removed, as shown in
Fig.~\ref{fig:US_FM_3CF_notCorrect} (c) and (d).

This observation suggests that the loss of precision is not accidental,
but is associated with the emergence of competing clustering
structures already visible in the unsupervised analysis.

\begin{figure}[th]
  \begin{center}
    \includegraphics[width=9cm]{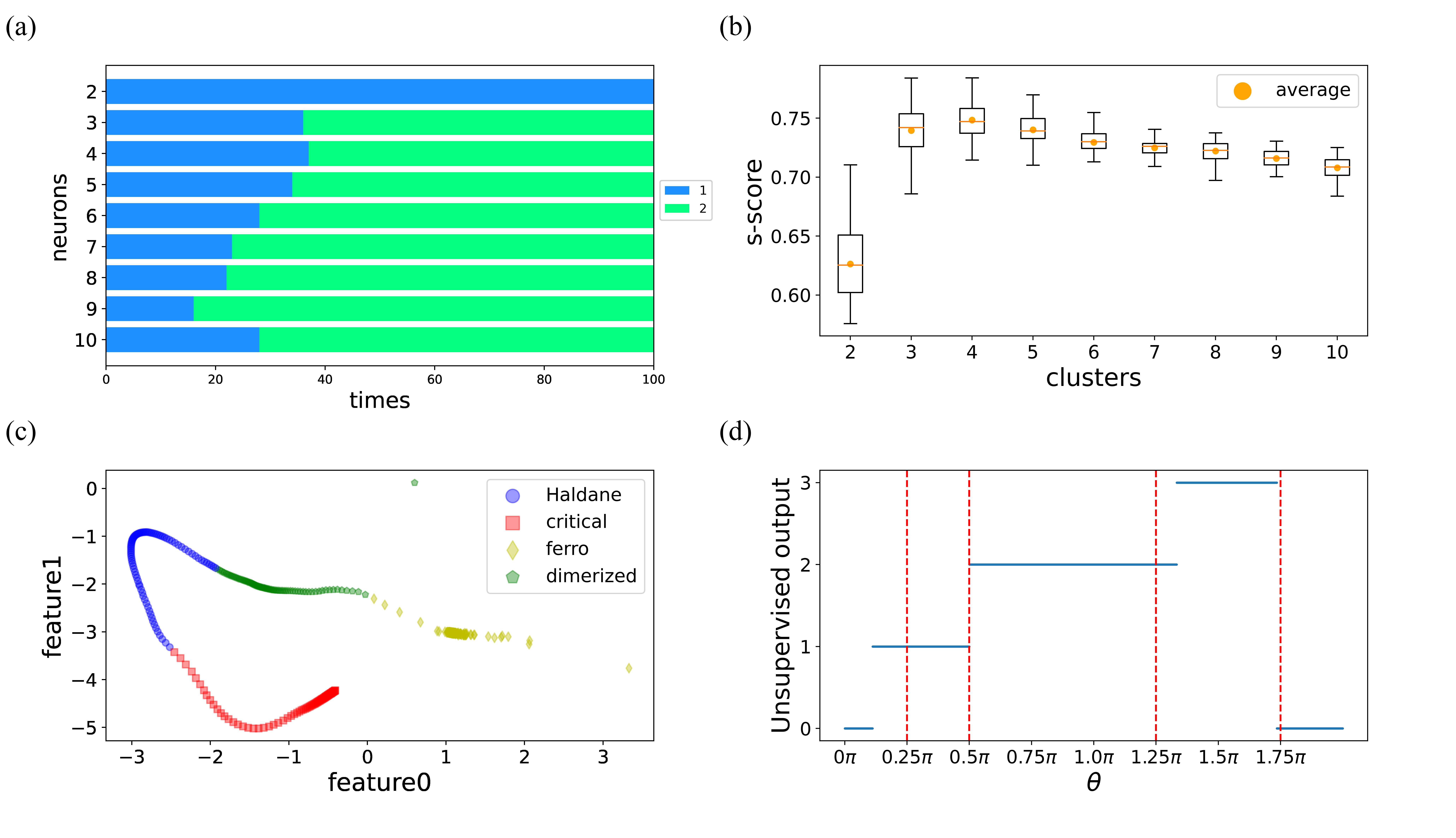}
    \caption{AE results using two correlation functions,
$C_{\mathrm{SS}}(r)$ and $C_{\mathrm{SU(3)}}(r)$. Panels (a), (b), (c), and (d) are the
same as those in Fig.~\ref{fig:AEResults_FMDMTMSU3}.
}
 \label{fig:AEResults_FMSU3}
  \end{center}
\end{figure}

\begin{figure}[th]
  \begin{center}
    \includegraphics[width=9cm]{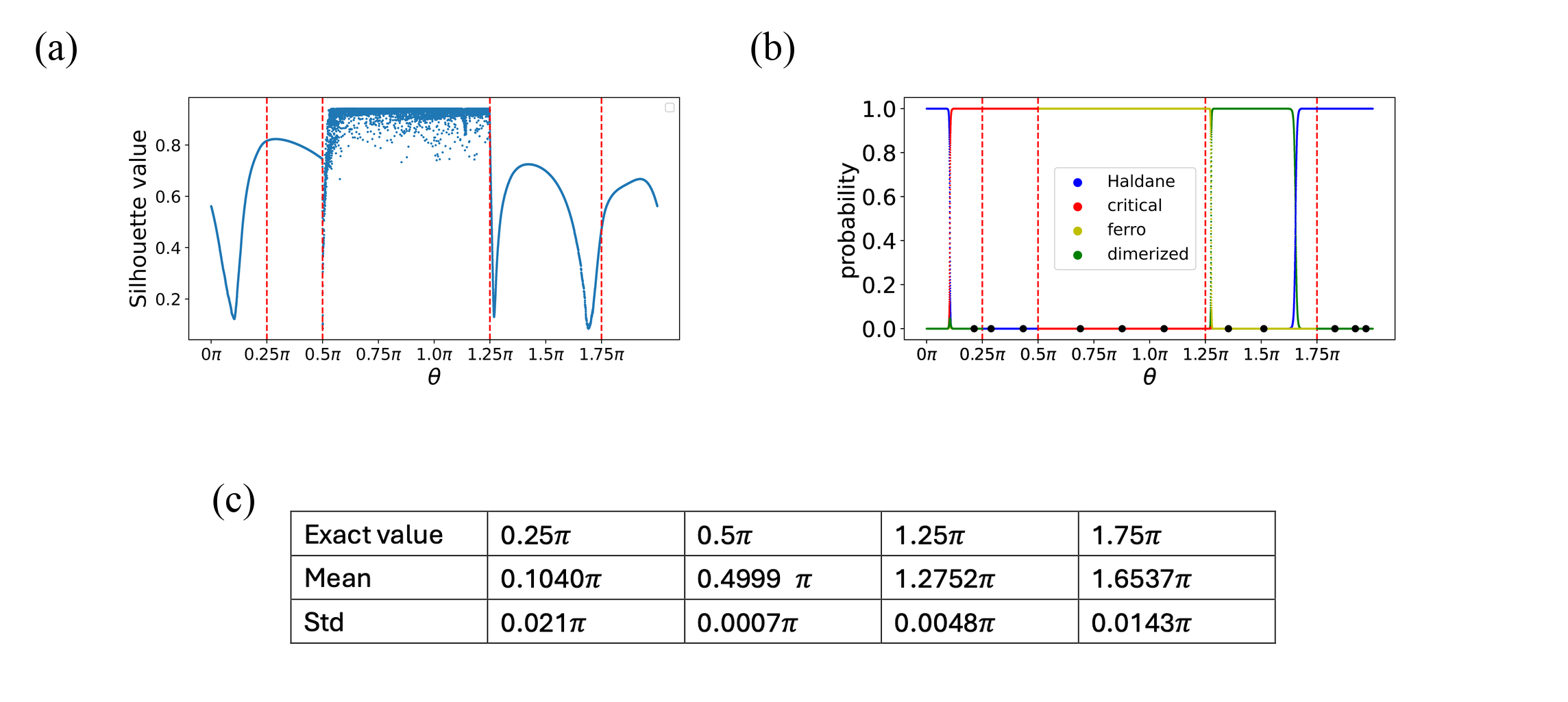}
    \caption{
Results obtained using two correlation functions,
$C_{\mathrm{SS}}(r)$ and $C_{\mathrm{SU(3)}}(r)$.
(a) Silhouette values. (b) Supervised-learning refinement based
on the high-confidence training points selected from the silhouette
values. The black dots indicate the selected training points, at
$\theta/\pi=0.2131$, $0.2890$, $0.4327$ (critical trimerized);
$0.6875$, $0.8750$, $1.0625$ (ferromagnetic); $1.3526$, $1.4222$,
$1.5117$ (dimerized); and $1.8302$, $1.9219$, $1.9691$ (Haldane). (c) Predicted phase-transition points.
}
    \label{fig:SCSV_FMSU3}
  \end{center}
\end{figure}

 We next consider another two-correlation combination,
$C_{\mathrm{SS}}(r)$ and $C_{\mathrm{SU(3)}}(r)$. Since these two
correlations are closely related to the underlying $SU(2)$ symmetry
and the emergent $SU(3)$ symmetry, respectively, we refer to this
combination as the symmetric correlations.

Fig.~\ref{fig:AEResults_FMSU3} shows the unsupervised results obtained
from these two correlations. As shown in
Fig.~\ref{fig:AEResults_FMSU3}(a) and (b), the latent dimension is
again $d_z=2$ and the silhouette analysis again selects $n=4$, so that
the correct number of phases is recovered from these two correlations
alone.

Fig.~\ref{fig:AEResults_FMSU3}(c) and (d) show the latent feature map
and the corresponding unsupervised phase assignments, respectively.
The results are slightly different from the other $n=4$ cases
discussed above. In particular, the unsupervised result suggests a
wider extension of the critical phase toward the Haldane phase, and
of the ferromagnetic phase toward the dimerized phase. This
behavior will be examined further using supervised learning. 

Fig.~\ref{fig:SCSV_FMSU3}(a) shows the SVs obtained
using $C_{\mathrm{SS}}(r)$ and $C_{\mathrm{SU(3)}}(r)$ as the input
data. The training points, selected as described in
Sec.~\ref{sec:protocol}, are marked by black dots in
Fig.~\ref{fig:SCSV_FMSU3}(b).

Interestingly, one of the selected training points for the critical
phase, chosen from the region where the SV remains above $95\%$ of
the maximum SV, lies inside the Haldane phase (at $\theta=0.2131\pi$). This is consistent with
the unsupervised result discussed above, where the critical phase
extends more broadly toward the Haldane phase.

The results of the supervised learning are shown in
Fig.~\ref{fig:SCSV_FMSU3}(b) and (c), where the predicted critical
points and their standard deviations over $100$ independent ML runs
are listed in Fig.~\ref{fig:SCSV_FMSU3}(c).

In this case, the predicted critical point between the critical and
ferromagnetic phases, $\theta_c=0.5\pi$, as well as that between the
ferromagnetic and dimerized phases, $\theta_c=1.25\pi$, remains in
good agreement with the exact values. However, the predicted critical
point between the Haldane and dimerized phases, $\theta_c=1.75\pi$,
shows a sizable deviation of about $0.1\pi$. The largest deviation
appears at the boundary between the Haldane and critical phases,
where the predicted critical point is $0.1040\pi$, differing from the
exact value $\theta_c=0.25\pi$ by about $0.15\pi$.

Using the two symmetric correlations shifts both boundaries of the
Haldane phase: the critical phase extends into the Haldane region on
one side, while the Haldane phase extends into the dimerized region on
the other. The former effect may be attributed to the $SU(3)$
correlation, and the latter to the absence of the dimer-dimer
correlation, which leaves the dimer order invisible near the
Takhtajan-Babujian point. Both effects are confirmed by the
single-correlation results in Sec.~\ref{sec:1corr}.

Signs of this reduced accuracy can already be observed in the SV curve
shown in Fig.~\ref{fig:SCSV_FMSU3}(a). The position of the SV minimum
between the Haldane and dimerized phases is farther from the exact
critical point than in the other $n=4$ cases discussed above.
Moreover, the position of the minimum between the critical and Haldane
phases deviates even more strongly from the exact critical point.
These features are consistent with the deviations of the predicted
critical points obtained from the supervised learning.

\begin{figure}[th]
  \begin{center}
    \includegraphics[width=9cm]{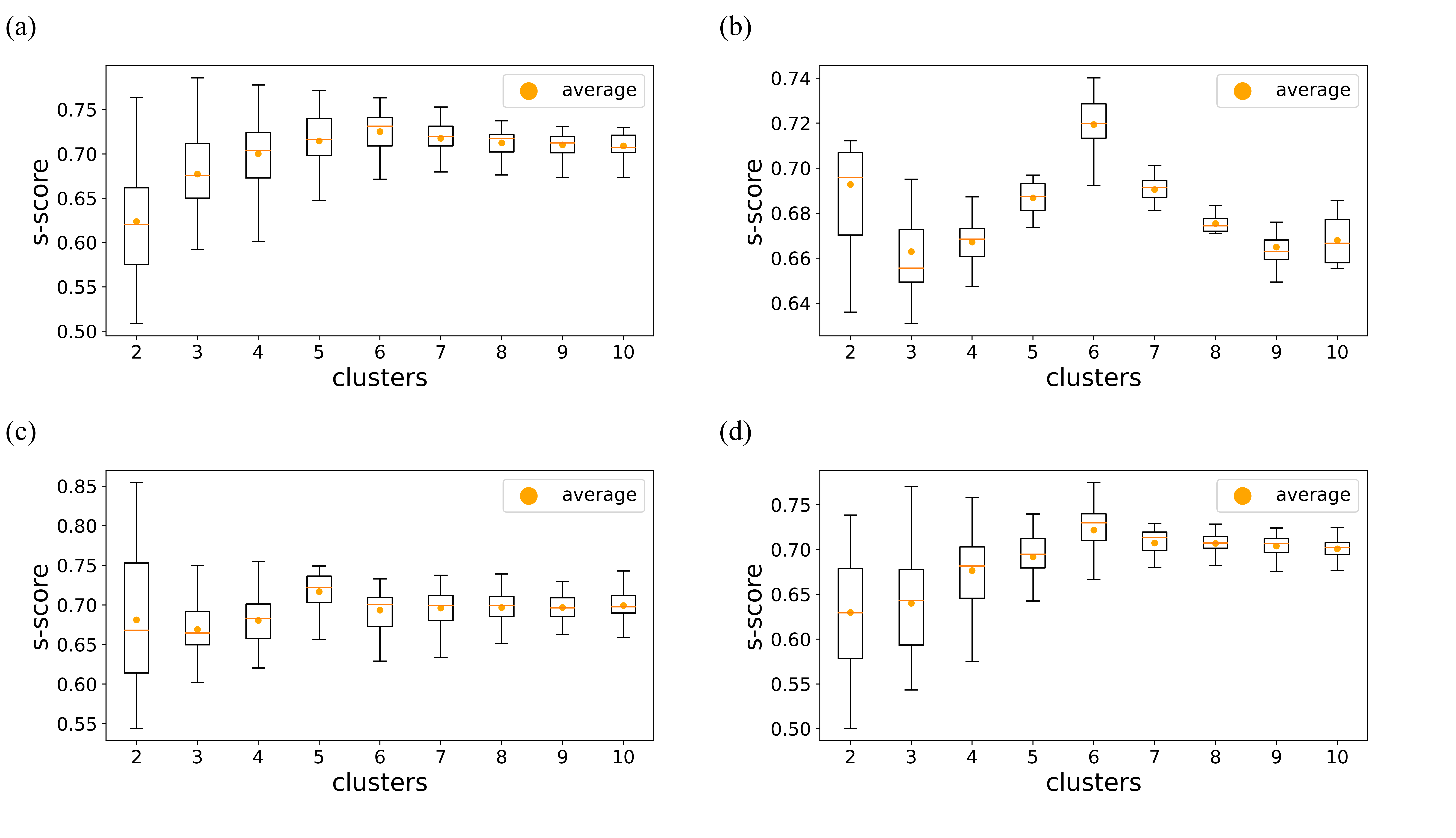}
    \caption{
    Silhouette scores for the two-correlation inputs with incorrect
    phase grouping.
    (a) $C_{\mathrm{DM}}(r)$ and $C_{\mathrm{SU(3)}}(r)$.
    (b) $C_{\mathrm{SS}}(r)$ and $C_{\mathrm{TM}}(r)$.
    (c) $C_{\mathrm{DM}}(r)$ and $C_{\mathrm{TM}}(r)$.
    (d) $C_{\mathrm{SU(3)}}(r)$ and $C_{\mathrm{TM}}(r)$.
    }
    \label{fig:SC_2CF_notCorrect}
  \end{center}
\end{figure}
\begin{figure}[th]
  \begin{center}
    \includegraphics[width=9cm]{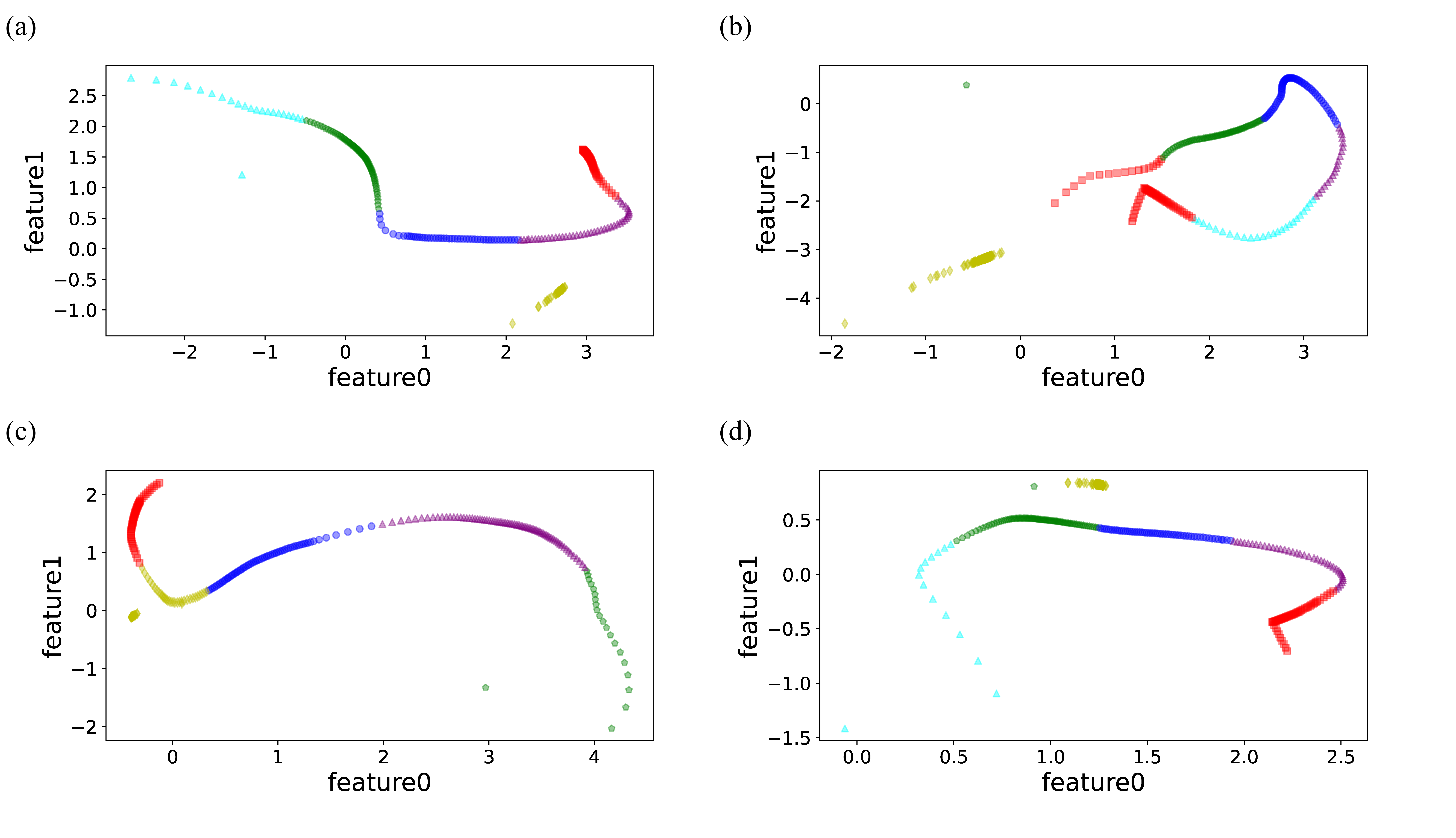}
    \caption{
    Latent feature maps for the two-correlation inputs with incorrect
    phase grouping.
    (a) $C_{\mathrm{DM}}(r)$ and $C_{\mathrm{SU(3)}}(r)$.
    (b) $C_{\mathrm{SS}}(r)$ and $C_{\mathrm{TM}}(r)$.
    (c) $C_{\mathrm{DM}}(r)$ and $C_{\mathrm{TM}}(r)$.
    (d) $C_{\mathrm{SU(3)}}(r)$ and $C_{\mathrm{TM}}(r)$.
    }
    \label{fig:FM_2CF_notCorrect}
  \end{center}
\end{figure}

The results for the other two-correlation combinations are not as good
as the previous cases. Fig.~\ref{fig:SC_2CF_notCorrect} shows the
silhouette scores for these remaining combinations. When
$C_{\mathrm{DM}}(r)$ and $C_{\mathrm{SU(3)}}(r)$,
$C_{\mathrm{SS}}(r)$ and $C_{\mathrm{TM}}(r)$, or
$C_{\mathrm{SU(3)}}(r)$ and $C_{\mathrm{TM}}(r)$ are used as input
data, as shown in Fig.~\ref{fig:SC_2CF_notCorrect}(a), (b), and
(d), respectively, the silhouette-score analysis suggests $n=6$ clusters. Only for
the case using $C_{\mathrm{DM}}(r)$ and $C_{\mathrm{TM}}(r)$, shown in
Fig.~\ref{fig:SC_2CF_notCorrect}(c), does the machine suggest
$n=5$ clusters.

Thus, none of these four two-correlation combinations gives the
correct four-phase grouping. Nevertheless, the incorrect groupings
still provide useful information about the role of each correlation
function in the phase-identification process.

Fig.~\ref{fig:FM_2CF_notCorrect} shows the latent feature maps
corresponding to these incorrect groupings, from which some
information can already be inferred. In panel (a), with
$C_{\mathrm{DM}}(r)$ and $C_{\mathrm{SU(3)}}(r)$ as the input data,
the three phases other than the ferromagnetic phase are separated into
five clusters along the continuous branch. In panel (b), with
$C_{\mathrm{SS}}(r)$ and $C_{\mathrm{TM}}(r)$, the three
non-ferromagnetic phases are again separated into five clusters;
moreover, two disconnected regions are assigned to the same cluster,
shown in red, even though they are separated in the feature map. In
panel (c), with $C_{\mathrm{DM}}(r)$ and $C_{\mathrm{TM}}(r)$, a
rather unusual situation occurs: the isolated ferromagnetic phase is
grouped together with a region lying on the continuous branch,
implying that part of the critical, dimerized, or Haldane phases is
incorrectly identified as belonging to the ferromagnetic phase. Since
K-means assigns each point to the nearest cluster center, the small
isolated ferromagnetic cluster absorbs the adjacent segment of the
continuous branch once the cluster number is incorrect. In panel (d),
with $C_{\mathrm{SU(3)}}(r)$ and $C_{\mathrm{TM}}(r)$, the dimerized,
critical, and Haldane phases are again separated into five clusters,
as in panel (a). These identifications can be seen more clearly from
the SVs and the unsupervised phase assignments discussed below.

\begin{figure}[th]
  \begin{center}
    \includegraphics[width=9cm]{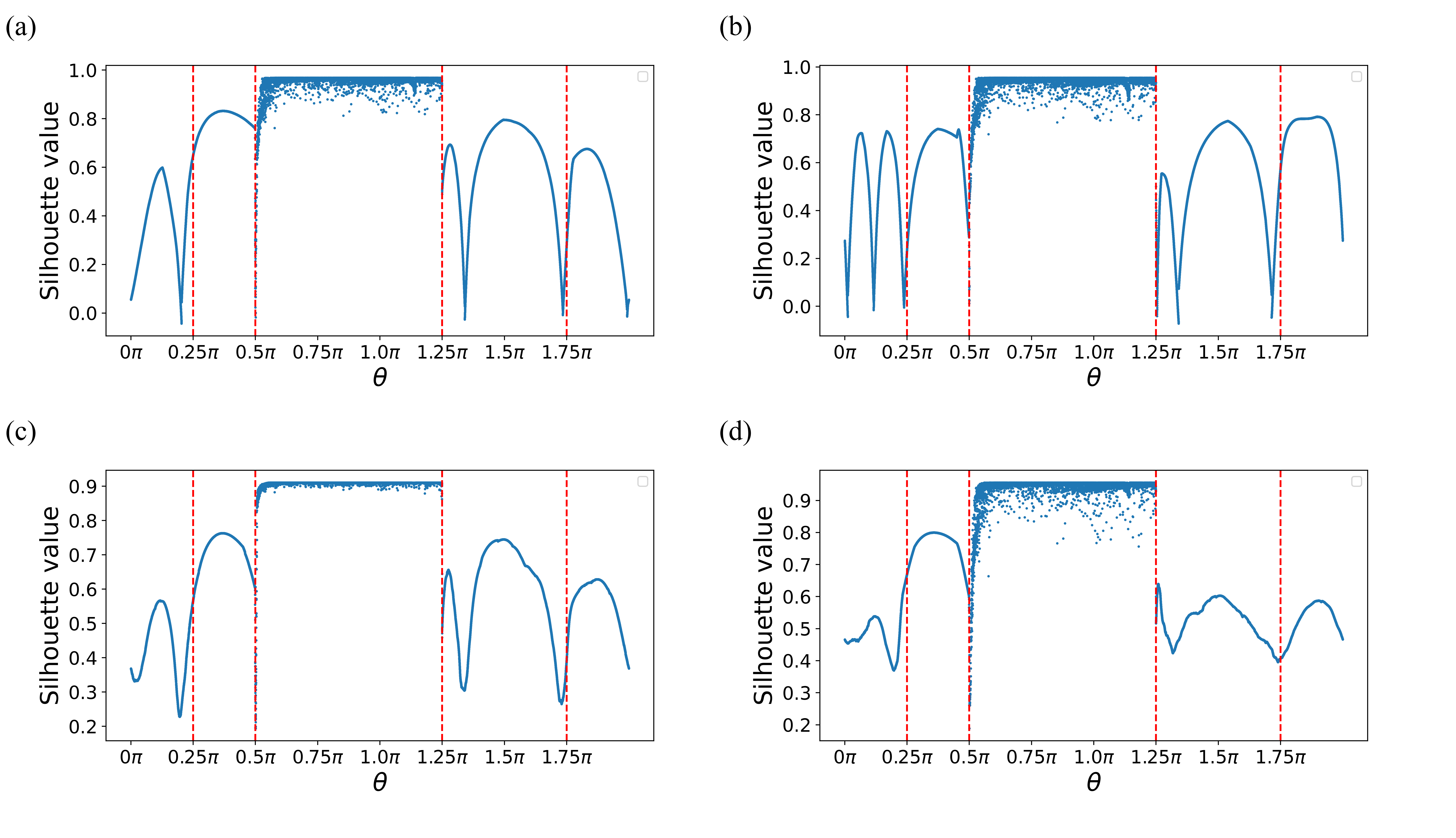}
    \caption{
    Silhouette values for the two-correlation inputs with incorrect
    phase grouping.
    (a) $C_{\mathrm{DM}}(r)$ and $C_{\mathrm{SU(3)}}(r)$.
    (b) $C_{\mathrm{SS}}(r)$ and $C_{\mathrm{TM}}(r)$.
    (c) $C_{\mathrm{DM}}(r)$ and $C_{\mathrm{TM}}(r)$.
    (d) $C_{\mathrm{SU(3)}}(r)$ and $C_{\mathrm{TM}}(r)$.
    }
    \label{fig:SV_2CF_notCorrect}
  \end{center}
\end{figure}

\begin{figure}[th]
  \begin{center}
    \includegraphics[width=9cm]{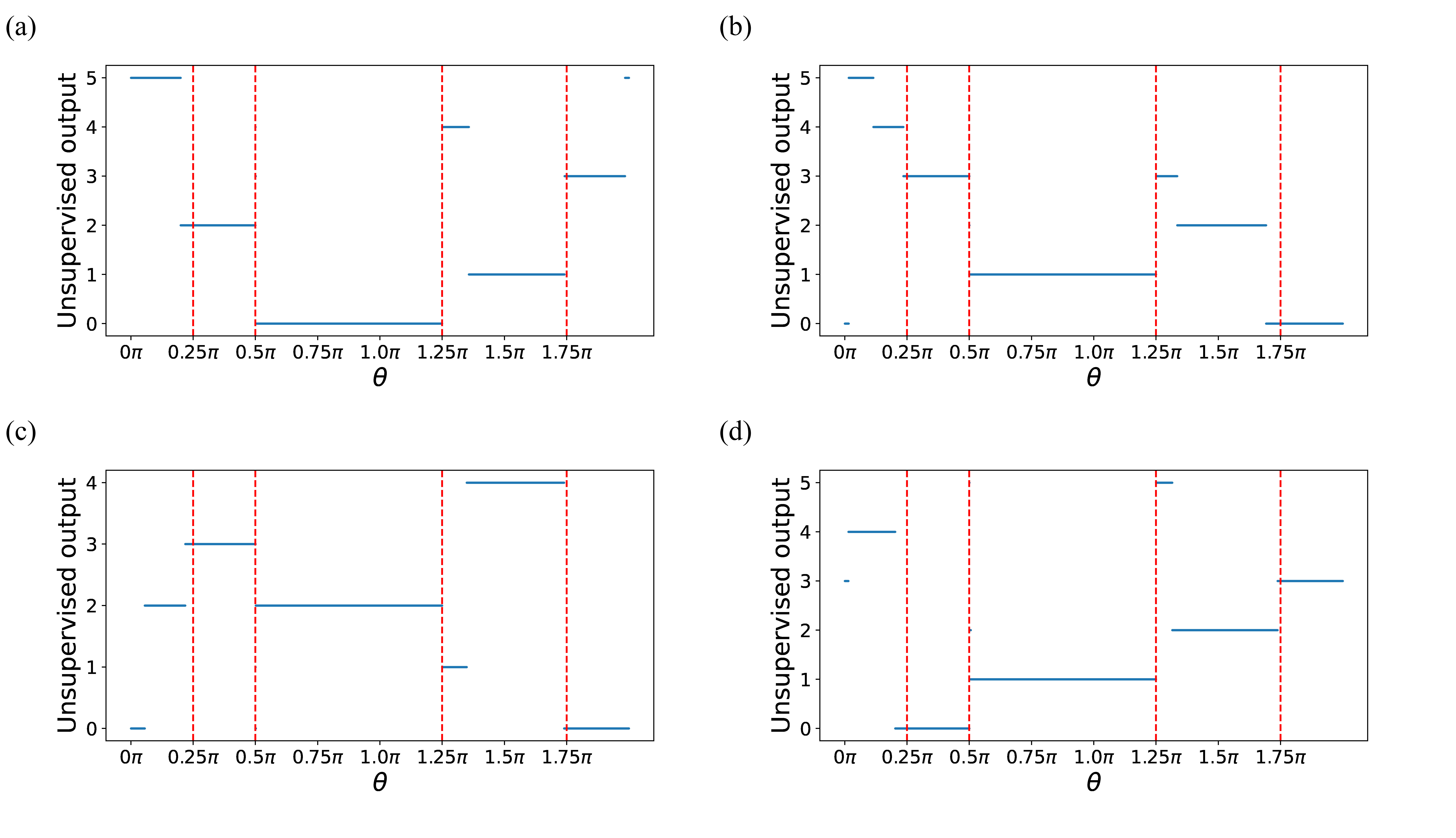}
    \caption{
    Unsupervised phase assignments for the two-correlation inputs with
incorrect phase grouping.
    (a) $C_{\mathrm{DM}}(r)$ and $C_{\mathrm{SU(3)}}(r)$.
    (b) $C_{\mathrm{SS}}(r)$ and $C_{\mathrm{TM}}(r)$.
    (c) $C_{\mathrm{DM}}(r)$ and $C_{\mathrm{TM}}(r)$.
    (d) $C_{\mathrm{SU(3)}}(r)$ and $C_{\mathrm{TM}}(r)$.
    The red dashed lines indicate the exact phase-transition points,
    and the cluster indices on the vertical axis are arbitrary.
    }
    \label{fig:USL_2CF_notCorrect}
  \end{center}
\end{figure}

The silhouette values and the unsupervised results for these incorrect
groupings are shown in Figs.~\ref{fig:SV_2CF_notCorrect} and
\ref{fig:USL_2CF_notCorrect}, respectively.

When $C_{\mathrm{DM}}(r)$ and $C_{\mathrm{SU(3)}}(r)$ are used as the
input data, as shown in Fig.~\ref{fig:SV_2CF_notCorrect}(a), the
Haldane and dimerized phases are each divided into two regimes by a
local minimum. Since the machine identifies six clusters, two
additional phases are expected to emerge from the Haldane and
dimerized phases. This expectation is confirmed by the unsupervised
result shown in Fig.~\ref{fig:USL_2CF_notCorrect}(a), where both the
Haldane and dimerized phases are explicitly split into two phases,
giving rise to two additional output phases.

When $C_{\mathrm{SS}}(r)$ and $C_{\mathrm{TM}}(r)$ are used, the SV
curve in Fig.~\ref{fig:SV_2CF_notCorrect}(b) shows seven minima: two
inside the Haldane phase and one inside the dimerized phase, in
addition to the four at the phase boundaries. Since the SV minima mark
the boundaries between neighboring clusters, seven boundaries for
$n=6$ clusters imply that one cluster consists of two disconnected
segments, as seen directly in the feature map of
Fig.~\ref{fig:FM_2CF_notCorrect}(b). The unsupervised assignment in
Fig.~\ref{fig:USL_2CF_notCorrect}(b) identifies this cluster: the
Haldane phase is split into three segments and the dimerized phase
into two, and the dimerized segment adjacent to the ferromagnetic
phase shares a cluster with the critical phase.

When $C_{\mathrm{DM}}(r)$ and $C_{\mathrm{TM}}(r)$ are used, the SV
curve in Fig.~\ref{fig:SV_2CF_notCorrect}(c) shows six minima, with
the Haldane and dimerized phases each split into two segments, while
only $n=5$ clusters are obtained. Again one cluster must be
disconnected, and Figs.~\ref{fig:FM_2CF_notCorrect}(c)
and~\ref{fig:USL_2CF_notCorrect}(c) show that it is the ferromagnetic
cluster, which absorbs one of the two Haldane segments.

Finally, when $C_{\mathrm{SU(3)}}(r)$ and $C_{\mathrm{TM}}(r)$ are
used, as shown in Fig.~\ref{fig:SV_2CF_notCorrect}(d), both the
Haldane and dimerized phases are divided into two regimes. Together
with the critical and ferromagnetic phases, this leads to six
clusters, consistent with the silhouette-score analysis. Therefore, the system is
again incorrectly separated into six clusters. This is confirmed by
the unsupervised result shown in
Fig.~\ref{fig:USL_2CF_notCorrect}(d).

We now summarize the results for the two-correlation inputs. When
$C_{\mathrm{SS}}(r)$ and $C_{\mathrm{DM}}(r)$ are used as the input
data, the machine correctly identifies four clusters, and the
predicted critical points are the most accurate among all
two-correlation cases. When $C_{\mathrm{SS}}(r)$ and
$C_{\mathrm{SU(3)}}(r)$ are used as the input data, the machine also
identifies four clusters, giving the second-best result. However, the
predicted critical points are less accurate than those obtained from
$C_{\mathrm{SS}}(r)$ and $C_{\mathrm{DM}}(r)$.

The other two-correlation combinations, on the other hand, cannot
correctly recover the four-phase grouping. Interestingly, the
ferromagnetic and critical phases are not further separated in any of
these cases, whereas the Haldane and dimerized phases are mostly split
into two or three different groups. A similar tendency will also be
observed in the one-correlation input results.

\subsubsection{1 Correlation}
\label{sec:1corr}

We now present the final set of results, where only one correlation
function is used as the input data. This analysis reveals how the
information carried by each individual correlation function affects
the identification of phases and the determination of critical
points.

\begin{figure}[th]
  \begin{center}
    \includegraphics[width=9cm]{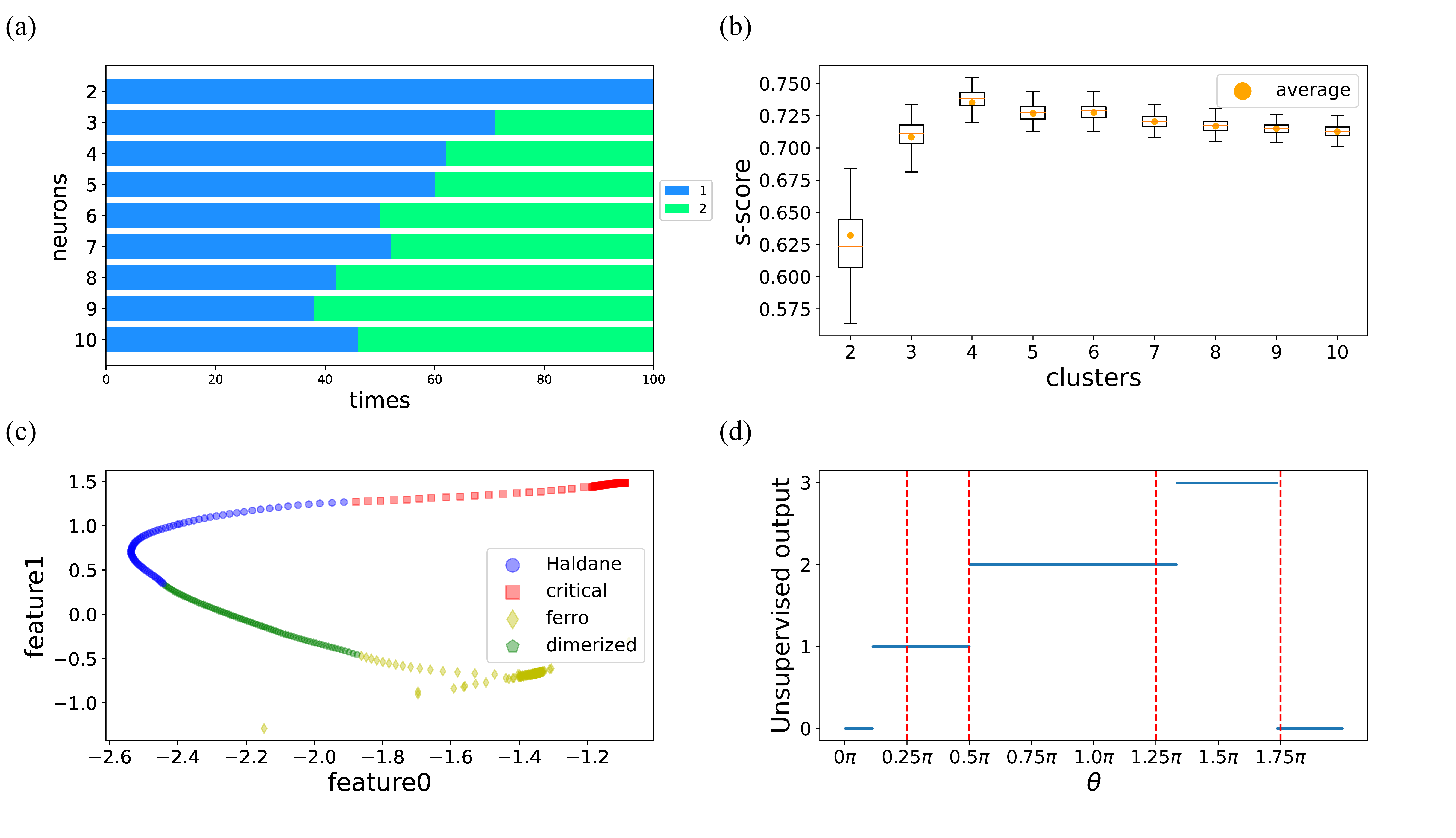}
    \caption{
    AE results using $C_{\mathrm{SU(3)}}(r)$ as the input data.
    Panels (a), (b), (c), and (d) are the same as those in
    Fig.~\ref{fig:AEResults_FMDMTMSU3}.
    }
    \label{fig:AEResults_SU3}
  \end{center}
\end{figure}

\begin{figure}[th]
  \begin{center}
    \includegraphics[width=9cm]{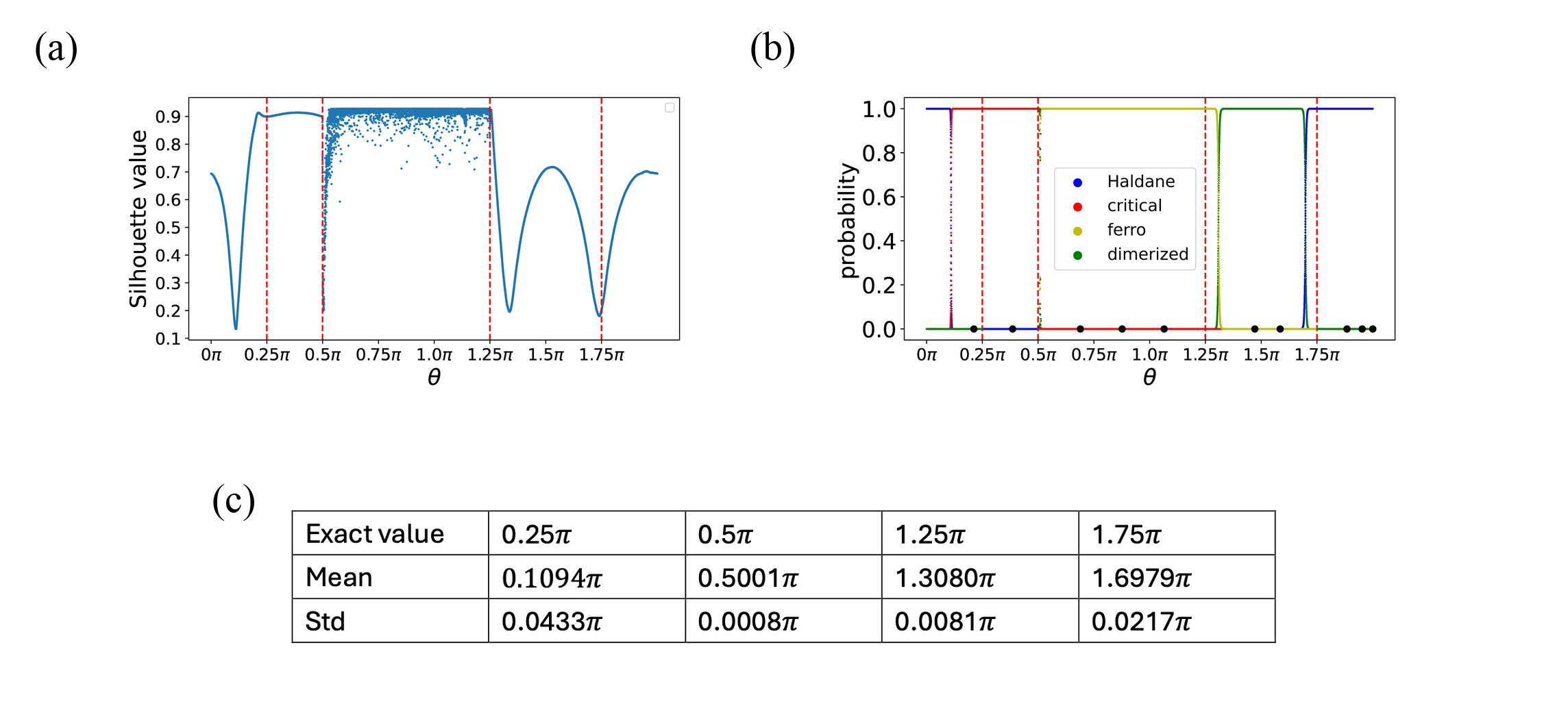}
    \caption{
    Results obtained using $C_{\mathrm{SU(3)}}(r)$ as the input data.
    (a) Silhouette values. (b) Supervised-learning refinement based
    on the high-confidence training points selected from the
    silhouette values. The black dots indicate the selected training
    points, at $\theta/\pi=0.2118$, $0.3856$ (critical trimerized);
    $0.6875$, $0.8750$, $1.0625$ (ferromagnetic); $1.4715$, $1.5297$,
    $1.5848$ (dimerized); and $1.8843$, $1.9523$, $1.9988$ (Haldane). (c) Predicted phase-transition points.
    }
    \label{fig:SCSV_SU3}
  \end{center}
\end{figure}

Fig.~\ref{fig:AEResults_SU3} shows the unsupervised results obtained
using $C_{\mathrm{SU(3)}}(r)$ as the input data. Surprisingly, the
silhouette analysis selects $n=4$ as the optimal cluster number in
this case. This means that the machine selects a four-cluster partition
that can be associated with the four phases using
only $C_{\mathrm{SU(3)}}(r)$, as shown in
Fig.~\ref{fig:AEResults_SU3}(b).

Fig.~\ref{fig:AEResults_SU3}(c) and (d) show the latent feature map
and the corresponding unsupervised phase assignments, respectively,
for a representative model whose behavior is shared by the majority
of the $100$ runs. The
results are less precise than those of the other four-cluster cases.
First, the critical regime is overestimated and extends significantly
into the Haldane phase. In addition, the ferromagnetic phase extends
into the dimerized regime.

The SVs for this case are shown in
Fig.~\ref{fig:SCSV_SU3}(a). One of the most interesting features is
that the SVs in the critical regime are very high, around $0.9$, and
nearly uniform, a behavior otherwise seen only in the ferromagnetic
phase. This may be understood from the renormalization-group
structure of the critical phase: for all $\pi/4<\theta<\pi/2$ the
marginal perturbation is irrelevant, so that the whole regime flows
toward the same $SU(3)_1$ fixed point~\cite{ItoiKato}. The $SU(3)$
symmetry, exact microscopically only at the ULS point, thus emerges at
long distances throughout the phase, and the $SU(3)$ correlation
matrices are very similar across the regime. These data points
therefore form a tight cluster in feature space, which yields
uniformly high SVs. The connection between the SVs and the RG
structure is discussed further in Sec.~\ref{sec:discussion}.

The training points are selected as described in
Sec.~\ref{sec:protocol}, except in the critical regime, where two
local maxima of the SV appear and both are chosen as training points;
they are marked by black dots in Fig.~\ref{fig:SCSV_SU3}(b).

The results of the supervised learning are shown in
Fig.~\ref{fig:SCSV_SU3}(b) and (c), where the predicted critical
points and their standard deviations over $100$ independent ML runs
are listed in Fig.~\ref{fig:SCSV_SU3}(c).

Because the SVs extend into the Haldane phase and one local maximum
appears directly inside the Haldane phase, the predicted critical
point between the critical and Haldane phases is shifted to
$0.1094\pi$. This gives a large deviation of about $0.141\pi$ from
the exact critical point at $0.25\pi$. The deviations of the other
critical points range from $0.0001\pi$ to $0.0580\pi$, which are much
less significant than the deviation found for the critical-Haldane transition.

\begin{figure}[th]
  \begin{center}
    \includegraphics[width=9cm]{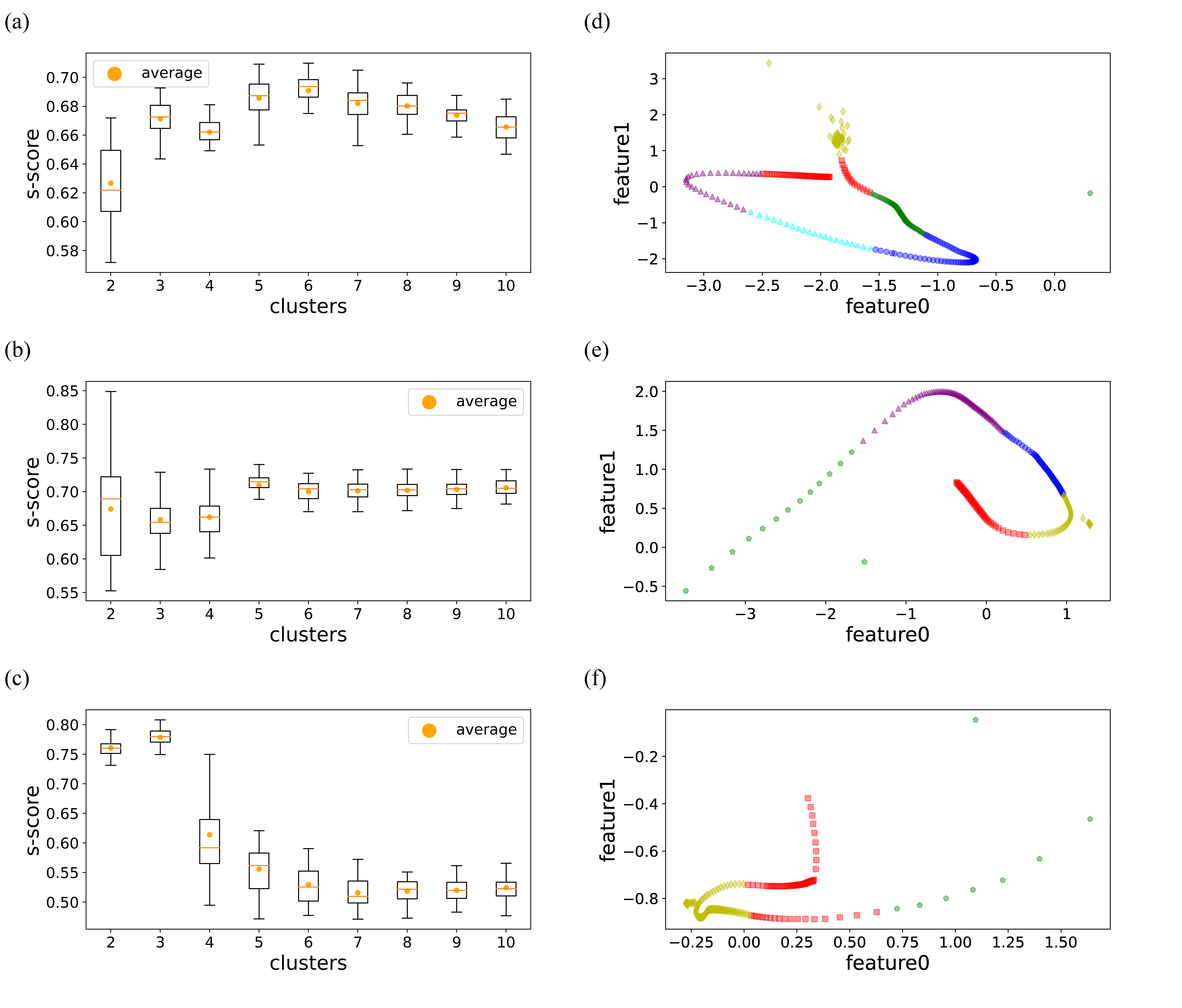}
    \caption{
    Silhouette scores and latent feature maps for the
    one-correlation inputs with incorrect phase grouping. Panels
    (a) and (d) correspond to $C_{\mathrm{SS}}(r)$,
    panels (b) and (e) correspond to $C_{\mathrm{DM}}(r)$,
    and panels (c) and (f) correspond to
    $C_{\mathrm{TM}}(r)$.
    }
    \label{fig:SC_FM_1CF_notCorrect}
  \end{center}
\end{figure}

\begin{figure}[th]
  \begin{center}
    \includegraphics[width=9cm]{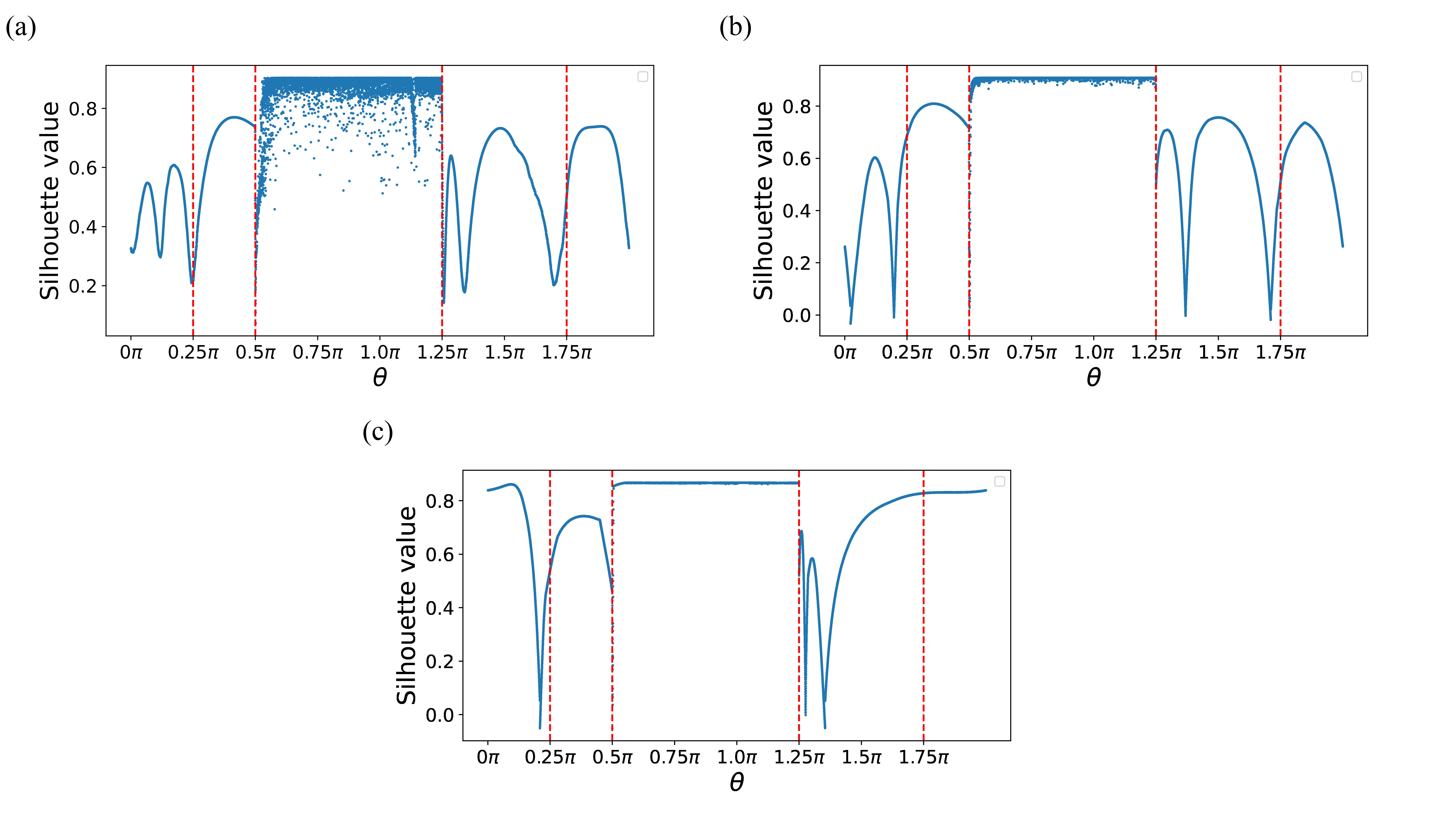}
    \caption{
    Silhouette values for the one-correlation inputs with incorrect
    phase grouping. Panel (a) corresponds to $C_{\mathrm{SS}}(r)$,
    panel (b) corresponds to $C_{\mathrm{DM}}(r)$, and panel (c)
    corresponds to $C_{\mathrm{TM}}(r)$.
    }
    \label{fig:SV_1CF_notCorrect}
  \end{center}
\end{figure}

\begin{figure}[th]
  \begin{center}
    \includegraphics[width=9cm]{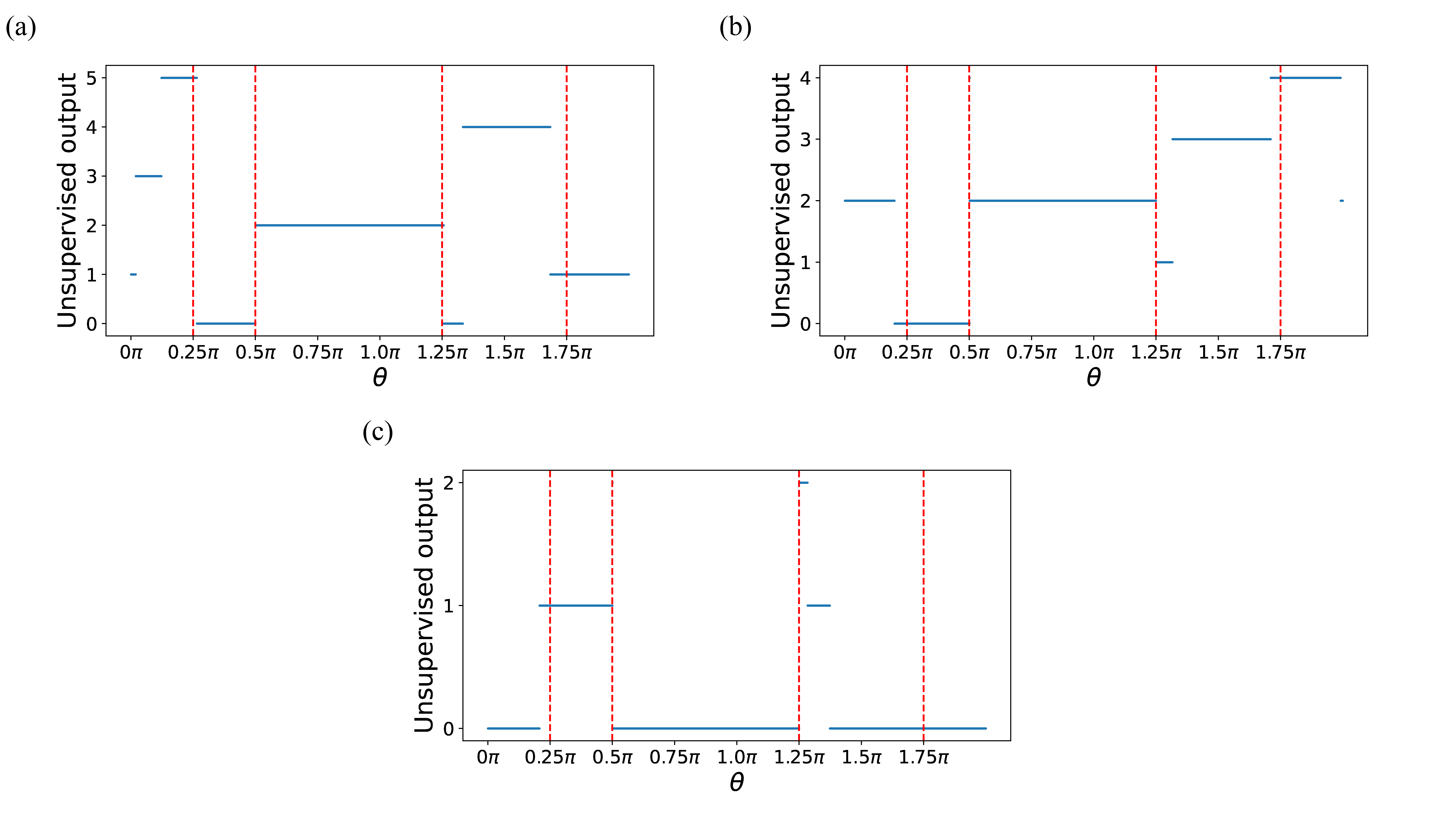}
    \caption{
    Unsupervised phase assignments obtained from the latent
    representations. Panels (a), (b), and (c) correspond to
    $C_{\mathrm{SS}}(r)$, $C_{\mathrm{DM}}(r)$, and
    $C_{\mathrm{TM}}(r)$, respectively. The red dashed lines indicate
    the exact phase-transition points, and the cluster indices on the
    vertical axis are arbitrary.
    }
    \label{fig:USCP_1CF_notCorrect}
  \end{center}
\end{figure}

Except for the $SU(3)$ correlation, the machine gives incorrect
results for the other three single-correlation inputs.
Figs.~\ref{fig:SC_FM_1CF_notCorrect},
\ref{fig:SV_1CF_notCorrect}, and
\ref{fig:USCP_1CF_notCorrect} show the silhouette scores and latent feature maps,
the SVs, and the unsupervised results,
respectively, for the other three single-correlations as inputs.

For $C_{\mathrm{SS}}(r)$ as the input data, the silhouette-score analysis gives
$n=6$ as the optimal cluster number on average, as shown in
Fig.~\ref{fig:SC_FM_1CF_notCorrect}(a). The feature map in
Fig.~\ref{fig:SC_FM_1CF_notCorrect}(d) shows that the critical regime
is grouped together with another regime labeled in red. In addition,
one phase is continuously separated into three different groups, although its identity cannot yet be determined from the
feature map alone

From the SVs in Fig.~\ref{fig:SV_1CF_notCorrect}(a), one can see two
local minima in the Haldane phase and one local minimum in the
dimerized phase. This indicates that the Haldane phase is separated
into three different regimes, while the critical phase is combined
with one part of the dimerized phase. The unsupervised grouping shown
in Fig.~\ref{fig:USCP_1CF_notCorrect}(a) confirms this analysis.

For $C_{\mathrm{DM}}(r)$ as the input data, the silhouette-score analysis gives
$n=5$ as the optimal cluster number on average, as shown in
Fig.~\ref{fig:SC_FM_1CF_notCorrect}(b). Interestingly, the feature map
suggests that the FM phase is merged with an additional separated
regime, as indicated by the yellow-green color in
Fig.~\ref{fig:SC_FM_1CF_notCorrect}(e).

The SVs show one local minimum in the Haldane phase and one local
minimum in the dimerized phase, as shown in Fig.~\ref{fig:SV_1CF_notCorrect}(b) . From the position of the yellow-green
regime in the feature map, which lies beside the critical phase, we
can now identify the regime grouped together with the ferromagnetic
phase as belonging to the Haldane phase. The additional phase then
appears in the dimerized phase, consistent with the result $n=5$.
This behavior is confirmed by the unsupervised result shown in
Fig.~\ref{fig:USCP_1CF_notCorrect}(b).

Finally, when the input data are $C_{\mathrm{TM}}(r)$, the situation
is very unusual. First, the silhouette-score analysis gives $n=3$, as shown in
Fig.~\ref{fig:SC_FM_1CF_notCorrect}(c). The feature map, shown in
Fig.~\ref{fig:SC_FM_1CF_notCorrect}(f), suggests that the
ferromagnetic phase is merged with another regime, labeled in
yellow-green, while the critical phase is also combined with an
additional separated regime, labeled in red. The SVs, shown in
Fig.~\ref{fig:SV_1CF_notCorrect}(c), indicate that the Haldane phase
extends into the dimerized phase, while the dimerized phase contains
two additional local minima beside this continuously extended regime.
Therefore, this extended regime is grouped together with the
ferromagnetic phase, and the critical phase is merged with one small
part of the separated dimerized phase. As a result, we obtain
$n=3$. This result is confirmed by the unsupervised result shown in
Fig.~\ref{fig:USCP_1CF_notCorrect}(c).

Before summarizing the single-correlation inputs, one phenomenon
should be discussed. In nearly all cases with incorrect grouping,
including the three- and two-correlation inputs discussed above, a
small region of the dimerized phase adjacent to the ferromagnetic
phase, $1.25\pi<\theta\lesssim1.32\pi$, is separated from the rest of
the dimerized phase. The correlation matrices in this region differ
noticeably from those in the rest of the dimerized phase, and the
pipeline separates it accordingly, although it does not correspond to
a distinct phase. A similar behavior occurs inside the Haldane phase;
possible connections of these subdivisions to known crossovers in the
correlation structure are discussed in Sec.~\ref{sec:discussion}.

Among the four single-correlation inputs, only
$C_{\mathrm{SU(3)}}(r)$ selects a four-cluster partition that can be
associated with the four phases.
However, the predicted critical points are not sufficiently precise.
The critical phase extends over a wide range into the Haldane phase,
and the Haldane-dimerized boundary is also noticeably shifted into the dimerized phase.

On the other hand, the other three correlations give completely
incorrect numbers of clusters. For the spin-spin and dimer-dimer
correlations, the main issue comes from the splitting of the Haldane
phase. With the spin-spin correlation, the Haldane phase is separated
into three pieces, whereas with the dimer-dimer correlation, it is
separated into two. Namely, both $C_{\mathrm{SS}}(r)$ and
$C_{\mathrm{DM}}(r)$ tend to split the Haldane phase.

The behavior of $C_{\mathrm{TM}}(r)$ is completely different from
that of the other correlations. Although it also produces splitting,
it largely merges the Haldane, dimerized, and ferromagnetic phases
into one phase. Therefore, the trimer-trimer correlation is the worst
correlation function to use alone, as it merges distinct phases and
predicts only three phases for the BLBQ model.

Table~\ref{tab:summary} summarizes the unsupervised results for all
fifteen combinations of correlation functions.

\begin{table*}[t]
\caption{Summary of the unsupervised results for all combinations of
correlation functions: selected cluster number $n$, predicted phase
boundaries $\theta_c/\pi$ from the supervised refinement (exact values
$0.25$, $0.5$, $1.25$, $1.75$) for the cases with $n=4$, and the
failure mode otherwise. Standard deviations are given in the
corresponding figures.}
\label{tab:summary}
\begin{ruledtabular}
\begin{tabular}{lcllll}
Input & $n$ & \multicolumn{4}{c}{$\theta_c/\pi$ or failure mode} \\
\hline
SS, DM, TM, SU(3) & 4 & 0.2488 & 0.5001 & 1.2583 & 1.7503 \\
SS, DM, SU(3) & 4 & 0.2472 & 0.5001 & 1.2624 & 1.7517 \\
SS, DM, TM & 5 & \multicolumn{4}{l}{extra cluster between FM and DM} \\
SS, TM, SU(3) & 3 & \multicolumn{4}{l}{Haldane merged with neighbors} \\
DM, TM, SU(3) & 6 & \multicolumn{4}{l}{extra clusters inside DM and Haldane} \\
SS, DM & 4 & 0.1762 & 0.5024 & 1.3326 & 1.7482 \\
SS, SU(3) & 4 & 0.1040 & 0.4999 & 1.2752 & 1.6537 \\
DM, SU(3) & 6 & \multicolumn{4}{l}{Haldane and DM each split in two} \\
SS, TM & 6 & \multicolumn{4}{l}{Haldane split in three, DM in two} \\
DM, TM & 5 & \multicolumn{4}{l}{part of Haldane merged with FM} \\
TM, SU(3) & 6 & \multicolumn{4}{l}{Haldane and DM each split in two} \\
SU(3) & 4 & 0.1094 & 0.5001 & 1.3080 & 1.6979 \\
SS & 6 & \multicolumn{4}{l}{Haldane split in three, critical merged with part of DM} \\
DM & 5 & \multicolumn{4}{l}{part of Haldane merged with FM, DM split} \\
TM & 3 & \multicolumn{4}{l}{FM, DM, and Haldane merged} \\
\end{tabular}
\end{ruledtabular}
\end{table*}

\section{Discussion}
\label{sec:discussion}
\subsection{The role of different correlation functions}

In fact, using a single correlation function as the input is not
sufficient to distinguish the four phases accurately, because the
phase structures of the BLBQ model are complex. Only the
$SU(3)$-correlation function can identify four phases, but the
predicted critical points are not accurate. In particular,
$C_{\mathrm{SU(3)}}(r)$ overextends the critical regime into the
Haldane phase. In contrast, the spin-spin and dimer-dimer correlations
tend to split the Haldane phase, while the trimer-trimer correlation
merges different phases together. Therefore, single-correlation inputs
generally perform poorly in predicting the critical points.

By using two different types of correlation functions, the situation
is improved. The best two-correlation combination is
$C_{\mathrm{SS}}(r)$ and $C_{\mathrm{DM}}(r)$, although this
combination still slightly extends the critical phase into the Haldane
phase and the ferromagnetic phase into the dimerized phase. The
second-best choice is the combination of $C_{\mathrm{SS}}(r)$ and
$C_{\mathrm{SU(3)}}(r)$. However, the result is very similar to that
obtained using the $SU(3)$ correlation alone. This indicates that the
spin-spin correlation does not significantly improve the performance
of the $SU(3)$ correlation, but it also does not degrade the
prediction. The other two-correlation combinations predict incorrect
numbers of phases.

By using $C_{\mathrm{SS}}(r)$, $C_{\mathrm{DM}}(r)$, and
$C_{\mathrm{SU(3)}}(r)$, we obtain critical points that are very close
to the exact values. The phase-extension problems are largely removed,
and the issue near the ferromagnetic-dimerized boundary also
disappears. When all four correlations are used, the results become
even more precise than those obtained from the three-correlation
combination without the trimer-trimer correlation.

The trimer-trimer correlation degrades the predictions
unless all four correlations are used simultaneously. As a
single-correlation input, it merges the ferromagnetic, dimerized, and
Haldane phases into a single phase. Moreover, when combined with one or two other
correlations, the predictions are mostly incorrect.

The Haldane phase is difficult to identify directly from the
local correlation functions considered here, because its conventional
diagnostic is the nonlocal string order parameter rather than a local order
parameter~\cite{denNijs1989,KennedyTasaki1992}. By using $C_{\mathrm{SU(3)}}(r)$, part of the
Haldane phase can be identified, although the predicted critical
points around it are not accurate. Single correlations, such as
$C_{\mathrm{SS}}(r)$ and $C_{\mathrm{DM}}(r)$, tend to split the
Haldane phase into several pieces. Interestingly, this behavior can be
corrected by combining these two correlations.

This observation may be understood from the information contained in
the correlation functions. Both $C_{\mathrm{SU(3)}}(r)$ and the
combination of $C_{\mathrm{SS}}(r)$ and $C_{\mathrm{DM}}(r)$ contain
two types of information: the linear spin-spin correlations and the
quadratic spin-spin correlations. The $SU(3)$ correlation contains
both contributions, but in a restricted manner since they appear only
through a fixed linear combination. Consequently, its predictive power
is limited. In contrast, the combination of the spin-spin and
dimer-dimer correlations effectively provides independent access to
both linear and quadratic spin-spin correlations, recovering most of
the information encoded in the BLBQ Hamiltonian. This may explain why
their combination predicts the phase boundaries much more accurately
than single-correlation inputs.

The correlation functions also admit a simple interpretation as
fluctuations of local Hamiltonian terms. Since
$D_i=\mathbf{S}_i\cdot\mathbf{S}_{i+1}$ is the bond Hamiltonian of the
Heisenberg chain, the variance of $\sum_i D_i$, i.e., the sum of the
connected dimer correlations over all pairs, vanishes at $\theta=0$
where the ground state is its eigenstate, although the individual raw
correlators in $C_{\mathrm{DM}}(r)$ do not; the nearest-neighbor element of $C_{\mathrm{SS}}(r)$ is
the same bond energy, and $C_{\mathrm{SS}}(r)$ as a whole is related
to the static structure factor. This viewpoint suggests that the
additional SV minima inside the Haldane phase, near
$\theta\approx0.12\pi$ for $C_{\mathrm{SS}}(r)$ and $\theta\approx0$ for
$C_{\mathrm{DM}}(r)$ [Fig.~\ref{fig:SV_1CF_notCorrect}(a) and (b)],
may reflect known crossovers in the correlation structure within the
Haldane phase rather than failures of the pipeline: the AKLT point
$\theta=\tan^{-1}(1/3)\approx0.10\pi$ is a disorder point at which
incommensurate spin correlations set in~\cite{Schollwock1996b}, and a
Lifshitz point, where the structure-factor maximum moves away from
$q=\pi$, lies slightly above it~\cite{Golinelli1999}. Whether the
subdivisions found by the pipeline track these crossovers is not
established here and is a promising direction for future work.

\subsection{Relations to the critical phenomena and renormalization
  group analysis}
\label{sec:Relations}

In the feature maps, the disconnected ferromagnetic cluster and the
connected remaining clusters are consistent with the discontinuous and
continuous nature of the respective transitions for the present
representation. This
behavior was also reported in our previous work on the XXZ
model~\cite{Ming23}, and in the BLBQ model the ferromagnetic phase is
likewise clearly disconnected from the other phases.

Interestingly, the SVs in the ferromagnetic phase always remain very 
high and nearly identical. This is particularly evident for the
spin-spin correlation input, shown in
Fig.~\ref{fig:SV_1CF_notCorrect}, since the spin-spin correlation is
the most relevant correlation for the ferromagnetic phase. Similar
behavior was also observed for the antiferromagnetic phase in the XXZ
model~\cite{Ming23}. This behavior has a simple origin. The fully
polarized ground state is independent of $\theta$, so that all
correlation matrices in the ferromagnetic phase are nearly
identical, the residual variation reflecting the long-wavelength
oscillations of the $SU(2)$-symmetric MPS approximation
(Sec.~\ref{sec:data}), and the
data points collapse onto a single cluster. In
renormalization-group language, the ferromagnetic phase is the basin
of a fully stable (sink) fixed point~\cite{Cardy} at which the system
already sits for every $\theta$, with vanishing correlation length;
the gapped Haldane and dimerized phases are likewise basins of sink
fixed points, but their finite correlation lengths vary with $\theta$
from well below to well above the block size, so that their
correlation data have not yet converged to the fixed-point form.

Similar behavior also occurs in the critical phase. As shown in
Fig.~\ref{fig:SCSV_SU3}, the SVs for $C_{\mathrm{SU(3)}}(r)$ are also
flat and nearly identical. According to the renormalization-group analysis, the whole critical
regime flows toward the $SU(3)$-symmetric ULS point, leading to an
emergent $SU(3)$ symmetry throughout the critical regime\cite{ItoiKato}. The
correlation matrices are therefore very similar throughout the phase,
up to the logarithmic corrections of the marginal operator, which is
reflected in the flat SV profile.

For the dimerized and Haldane phases, in contrast, the SVs form
concave curves with a maximum inside the phase when the number of
clusters is correctly identified. In these gapped phases the
correlation length varies from well below to well above the block size
$l_s$ as $\theta$ is varied, so that the correlation matrices retain a
visible $\theta$ dependence and become increasingly similar to those
of the neighboring phase toward the boundaries. Taken together, these
observations suggest that the SV profile of a phase reflects how
rapidly its correlation data converge, within the accessible block
size, to the form dictated by the fixed point governing the phase:
phases whose long-distance physics is $\theta$-independent, exactly or
up to marginal corrections, show flat profiles, whereas phases with a
$\theta$-dependent correlation length show concave ones. Establishing
a direct correspondence between silhouette profiles and the
renormalization-group structure would require testing the robustness
of these patterns under changes of block size, normalization, and
clustering method, which we leave for future work.

Finally, we discuss the importance of the four
correlation functions. $C_{\mathrm{SS}}(r)$ is the relevant
correlation for the ferromagnetic phase,
$C_{\mathrm{DM}}(r)$ is the relevant correlation for the
dimerized phase, and $C_{\mathrm{SU(3)}}(r)$ is the relevant
correlation for the critical (trimerized) phase.

The role of $C_{\mathrm{TM}}(r)$, however, is more subtle.
Although this phase is usually referred to as the critical
trimerized phase, the renormalization-group trajectories do not flow
toward a trimerized fixed point. Instead, they flow toward the
$SU(3)$-symmetric ULS point. Consequently, the phase is often
called the trimer-liquid phase in the recent
literature~\cite{Oh2017,MashikoNomura2021,MashikoNomura2023}.
Since $C_{\mathrm{TM}}(r)$ probes the short-range period-$3$
structure rather than the long-distance $SU(3)_1$ behavior, it need
not be a relevant correlation for phase identification, and our
results confirm that it is the least important one. By itself, it
cannot distinguish different phases and instead tends to merge
several phases together.

However, once the other correlation functions have already
identified the correct phases, the trimer-trimer correlation
still improves the precision of the predicted critical points,
as shown in
Figs.~\ref{fig:AEResults_FMDMTMSU3} and
\ref{fig:SCSV_FMDMTMSU3}. Therefore, by analogy with the
terminology of the renormalization group, we refer to
$C_{\mathrm{TM}}(r)$ as a \emph{marginal correlation}, in an empirical
sense. It does
not determine the phases themselves, but it refines the
predictions once the phases have already been identified.

\section{Conclusion and Outlook}
\label{sec:conclusion}

The BLBQ model has a complex and rich phase diagram. It contains a
topological Haldane phase, a critical phase with emergent symmetry, a
dimerized phase, and a ferromagnetic phase, together with two
discontinuous and two continuous phase transitions. In this work, we
study the BLBQ model using our machine-learning pipeline and reveal
how different correlation functions encode the information needed for
phase identification.

Four correlation functions, $C_{\mathrm{SS}}(r)$,
$C_{\mathrm{DM}}(r)$, $C_{\mathrm{SU(3)}}(r)$, and
$C_{\mathrm{TM}}(r)$, are used as input data to distinguish the
phases. We find that $C_{\mathrm{SU(3)}}(r)$ alone selects a four-cluster
partition that can be associated with the four phases, although some phase boundaries show large deviations.
$C_{\mathrm{SS}}(r)$ and $C_{\mathrm{DM}}(r)$ have to be used
together to recover the four phases. These three correlations
already identify the phase boundaries with good precision;
$C_{\mathrm{SS}}(r)$, $C_{\mathrm{DM}}(r)$, and $C_{\mathrm{SU(3)}}(r)$
are the relevant correlations for the ferromagnetic, dimerized, and
critical phases, respectively.

On the other hand, $C_{\mathrm{TM}}(r)$ cannot be used alone to
determine the critical points, because it tends to merge different
phases. However, when it is added to the other three
correlations, it refines the precision of the phase boundaries.
It is therefore referred to as a marginal correlation.

The latent representations and silhouette values reveal signatures
that we relate to the $\theta$ dependence of the correlations within
each phase. The
feature maps show disconnected and connected clusters consistent with
the discontinuous and continuous transitions, respectively. The SVs
are high and nearly flat in the ferromagnetic and critical phases,
whose long-distance physics is $\theta$-independent, and concave in
the gapped Haldane and dimerized phases, whose correlation length
varies with $\theta$. Whether this correspondence between silhouette
profiles and the renormalization-group structure of the phases holds
more generally is an intriguing question that we leave for future
work.

At this point, many interesting questions remain open. One direction
is to investigate other quantum phase transitions involving
topological phases and emergent symmetries, such as spin-$2$ quantum
spin models\cite{MingYip12,MingYip15}, emergent symmetries at deconfined quantum critical
points~\cite{DCPT}, and topological finite-temperature effects. Another,
potentially more fundamental direction is to explore the mathematical
structure of machine-learning pipelines and their connections to
renormalization-group theory and conformal field theory. We hope that the present work will motivate further investigations
along these directions.

\section{Acknowledgments}
M.-C.C. acknowledges NSTC support under Contract
No. 115-2112-M-005-005 and the Asian Office of
Aerospace Research and Development (AOARD) for support
under Award No. FA2386-23-1-4104.

\appendix
\section{Model Architectures}
\begin{table}
\caption{Autoencoder architecture}
\centering
\begin{tabular}{lll}
\hline\hline
Layer & Parameters & Activation \\
\hline
 & Input: $c \times 10 \times 10$ & \\
Conv. & $16 \times 3 \times 3$ & ReLU \\
Linear & $1600 \times 512$ & ReLU \\
Linear & $512 \times 256$ & ReLU \\
Linear & $256 \times 128$ & ReLU \\
Linear & $128 \times 64$ & ReLU \\
Linear & $64 \times \mathrm{features}$ & \\
Linear & $\mathrm{features} \times 64$ & ReLU \\
Linear & $64 \times 128$ & ReLU \\
Linear & $128 \times 256$ & ReLU \\
Linear & $256 \times 512$ & ReLU \\
Linear & $512 \times 1600$ & ReLU \\
Transposed conv. & $c \times 3 \times 3$ & Tanh \\
\hline\hline
\label{tab:architecture}
\end{tabular}
\end{table}

\begin{figure}[th]
  \begin{center}
    \includegraphics[width=9cm]{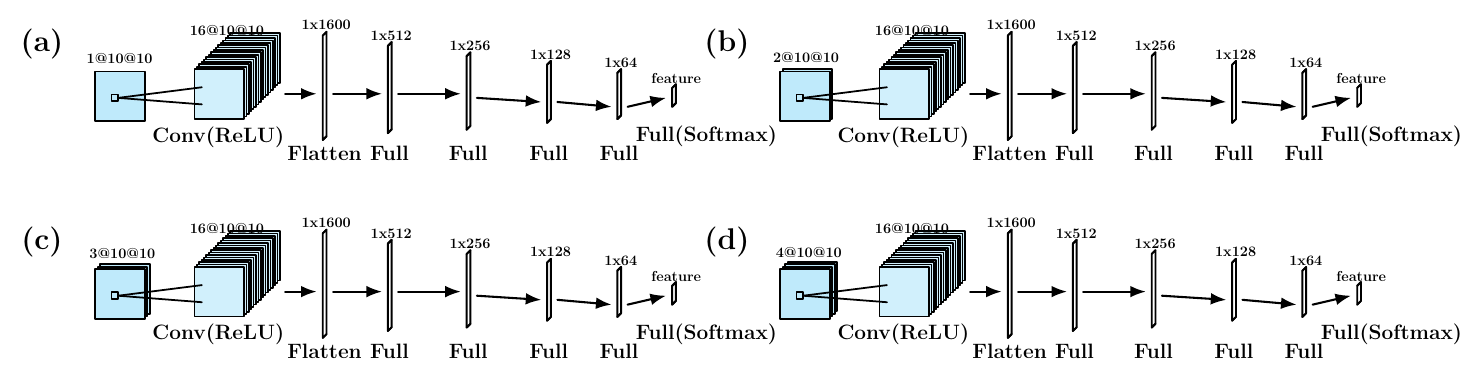}
    \caption{Convolutional neural networks used for the supervised
refinement, for (a) one, (b) two, (c) three, and (d) four input
channels.}
    \label{fig:CNN}
  \end{center}
\end{figure}

Here we present the neural-network architectures used for the
results in Sec.~\ref{sec:unsupervised}. Table~\ref{tab:architecture}
lists the autoencoder used for all unsupervised results
(Figs.~\ref{fig:AEResults_FMDMTMSU3}--\ref{fig:USCP_1CF_notCorrect}),
where $c$ is the number of input channels, i.e., the number of
correlation matrices used, and the width of the middlemost layer,
labeled ``features'', is $n_{\mathrm{mid}}$ in the main text. All
intermediate layers use rectified linear units
(ReLU)~\cite{NairHinton2010}, and the output layer of the decoder uses
the hyperbolic tangent. The classifier used for the supervised
refinement is the convolutional neural network shown in
Fig.~\ref{fig:CNN} for $c=1$ to $4$ input channels; its output layer
has $n$ neurons with a softmax activation, where $n$ is the number of
clusters determined by the K-means analysis.

The networks are implemented in Keras with the TensorFlow
backend~\cite{Keras,TensorFlow}, and PCA, K-means, and silhouette
analysis use scikit-learn~\cite{sklearn}. The autoencoder is trained
on all $20001$ images with the mean-squared reconstruction error,
using stochastic gradient descent with learning rate $10^{-4}$ and
batch size $16$, for up to $50$ epochs with early stopping (patience
$25$ epochs) monitored on the training reconstruction loss; no dropout
or weight decay is used. The classifier is trained with the
cross-entropy loss, using Adam with learning rate $10^{-4}$, batch
size $16$, and $50$ epochs, with $20\%$ of the labeled samples held
out for validation. The probability curves in
Fig.~\ref{fig:SCSV_FMDMTMSU3}(b) and the corresponding panels of later
figures are averages of the softmax outputs over the $100$ trained
models, whereas the mean and standard deviation of each critical point
are computed from the crossing points of the individual models.

\end{document}